\documentclass{aastex631}

\graphicspath{{./}{}}

\usepackage{amssymb}
\usepackage{amsmath}
\usepackage{cases}

\begin{document}

\title{Coronal rain in randomly heated arcades}


\author[0000-0001-8164-5633]{Xiaohong Li}
\affiliation{Centre for mathematical Plasma Astrophysics, Department of Mathematics, KU Leuven, \\
Celestijnenlaan 200B, 3001, Leuven, Belgium}

\author[0000-0003-3544-2733]{Rony Keppens}
\affiliation{Centre for mathematical Plasma Astrophysics, Department of Mathematics, KU Leuven, \\
Celestijnenlaan 200B, 3001, Leuven, Belgium}

\author[0000-0002-4391-393X]{Yuhao Zhou}
\affiliation{Centre for mathematical Plasma Astrophysics, Department of Mathematics, KU Leuven, \\
Celestijnenlaan 200B, 3001, Leuven, Belgium}

\begin{abstract}
Adopting the MPI-AMRVAC code, we present a 2.5-dimensional magnetohydrodynamic (MHD) simulation, which includes thermal conduction and radiative cooling, to investigate the formation and evolution of the coronal rain phenomenon. We perform the simulation in initially linear force-free magnetic fields which host chromospheric, transition region, and coronal plasma, with turbulent heating localized on their footpoints. Due to thermal instability, condensations start to occur at the loop top, and rebound shocks are generated by the siphon inflows. Condensations fragment into smaller blobs moving downwards and as they hit the lower atmosphere, concurrent upflows are triggered. Larger clumps show us clear ``coronal rain showers" as dark structures in synthetic EUV hot channels and bright blobs with cool cores in the 304 {\AA} channel, well resembling real observations. Following coronal rain dynamics for more than 10 hours, we carry out a statistical study of all coronal rain blobs to quantify their widths, lengths, areas, velocity distributions, and other properties. The coronal rain shows us continuous heating-condensation cycles, as well as cycles in EUV emissions. Compared to previous studies adopting steady heating, the rain happens faster and in more erratic cycles. Although most blobs are falling downward, upward-moving blobs exist at basically every moment. We also track the movement of individual blobs to study their dynamics and the forces driving their movements. The blobs have a prominence-corona transition-region-like structure surrounding them, and their movements are dominated by the pressure evolution in the very dynamic loop system.

\end{abstract}


\section{Introduction} \label{sec:intro}

Coronal rain is a widely observed phenomenon where dense and cool condensations form in the hot corona, and then fall down along magnetic loops to the solar surface \citep{Antolin2012}. It was first reported around 50 years ago \citep{Kawaguchi1970, Leroy1972}, and appears frequently in non-flaring coronal loops \citep{Schrijver2001, Antolin2020} and post-flare loops \citep{Scullion2016, Ruan2021}. Coronal rain has been observed off-limb and on-disk in chromospheric lines (e.g., Ca II H and H$\alpha$) and transition region (TR) lines (e.g., Si IV), as well as in extreme ultraviolet (EUV) observations \citep{Schrijver2001, Patrick2012, Verwichte2017, Li2020, Ishikawa2020}. It presents us with a clear multi-thermal coronal plasma, which displays the emission of different wavelengths co-spatially \citep{Antolin2015}. Coronal rain is measured to have a density of $\sim$ 10$^{10}$ $-$ 10$^{11}$ cm$^{-3}$ and falls down to the solar surface at velocities of 30 $-$ 200 km s$^{-1}$, with a mean value of $\sim$ 70 km s$^{-1}$ \citep{Schrijver2001, Kleint2014}. Coronal rain takes an essential part in the mass cycle between the chromosphere and the corona \citep{Li2018, Li2019, Li2020, Li2021a, Li2021b, Kohutova2019}.

Thermal instability has been accepted to be an important mechanism for the formation of condensations \citep{Parker1953, Field1965, Claes2019, Claes2020}. Uniform coronal plasmas can be regarded as in thermal equilibrium when its density and temperature provide a perfect balance between heating and cooling. Such equilibrium can be easily destroyed with a small initial perturbation \citep{Klimchuk2019}. The heating input of a coronal loop is usually considered to be localized at its footpoints \citep{Antiochos1999, Aschwanden2001, Karpen2001, Muller2003, Muller2004, Muller2005, Xia2011}, which causes the evaporation of chromospheric plasma, and therefore the loop will become denser and hotter. When energy losses caused by radiation exceed the heating input, the temperature decreases over time. Since the radiation loss function is negatively related to temperature in the coronal temperature range, the decrease in temperature will further increase the radiation, resulting in catastrophic cooling. Then gas pressure and temperature in the perturbed region decline rapidly, which leads to ambient plasma being inhaled to form cooler and larger condensations. This out-of-control effect will continue to reduce the temperature and increase the density until cooling and heating reach equilibrium again. Modern views also allow for the distinct possibility that such a thermal equilibrium is actually never achieved, in so-called thermal non-equilibrium cycles \citep{Antolin2020}. Condensations will appear as either coronal rain or prominences/filaments affected by the local magnetic field configuration. The condensations formed in arched loop may fall down along the loop legs in a short time and appear as coronal rain, while the condensations in a prominence may remain suspended for many hours to several days hosted by magnetic dips \citep[e.g.][]{Antiochos1994, Jenkins2021, Adrover2021}. 

This chromospheric evaporation$-$coronal condensation scenario was first demonstrated in one-dimensional (1D) hydrodynamic simulations along individual magnetic loops \citep{Antiochos1999} and then widely used in different simulations to explain the occurrence and dynamics of prominences and coronal rain \citep{Karpen2001, Karpen2006, Antolin2010, Xia2011, Luna2012, Froment2018, Gabriel2021}. Comparing the results from 1.5D magnetohydrodynamic (MHD) simulations of coronal loops heated by different mechanisms such as footpoint nanoflare heating and Alfvén waves heating with the observations, \citet{Antolin2010} noticed that coronal rain is inhibited when the coronal loops are primarily heated by Alfvén waves and proposed that it can well reflect the properties of the coronal heating mechanisms. \citet{Fang2013} presented the first 2.5D MHD simulations of coronal rain. In their simulations, they studied in detail how large zigzag shape condensations formed at the top regions of magnetic loops, which later fell down and split into fragments. They statistically analyzed the widths, lengths, velocities, and other characteristics of the coronal rain blobs that appeared within 80 minutes of physical time. By extending their simulation with a higher resolution and much longer running time, \citet{Fang2015} further studied the formation of coronal rain condensations and found that coronal rain showers can occur in limit cycles. They analyzed rebound shocks and the prominence-corona transition-region (PCTR) structures around the blobs and explained the mechanisms leading to the generation of concurrent upflows and counter-streaming flows. \citet{Moschou2015} simulated coronal rain phenomenon in 3D settings of a quadrupolar arcade magnetic configuration, in which coronal rain appeared continuously and exhibited obvious features of Rayleigh$-$Taylor or interchange instability. \citet{Xia2017} performed a 3D MHD simulation on coronal rain in a weak bipolar magnetic field, a typical coronal magnetic topology where coronal rain occurs. They analyzed the characteristics of the coronal rain blobs statistically and found that the majority of the blobs have masses less than 10$^{10}$ g. \citet{Kohutova2020} realized a self-consistent 3D simulation about the coronal rain formation, in which chromospheric evaporation is produced by Ohmic and viscous heating through the braiding of the magnetic field due to magnetoconvection, and then drives thermal instability in the corona and leads to condensation formation.

These numerical simulations based on the evaporation$-$condensation model with various setups have enriched our understanding on the mechanism and dynamic of condensations, and mass transfer and energy conversion in the solar atmosphere. Nearly all of them have so far relied on artificially static and localized heating functions added to the energy equation. The localized heating functions adopted in previous simulation work are generally steady and uniform \citep[e.g.][]{Fang2013, Fang2015, Xia2017}. Since moving magnetic features and turbulent convection \citep{Tu2013} are almost ubiquitous in the solar lower atmosphere, the quasi-constant heating profiles are too simplistic for the coronal rain simulation. Coronal heating should be an impulsive phenomenon, affected by multi-scale and continuous disturbances in the solar photosphere. \citet{Zhou2020} proposed that the localized heating should be turbulent, which is randomly distributed, and they successfully simulate the fine structure and the counterstreaming flows of solar filaments by investigating the response of the coronal loops to randomized heating imposed toward the footpoints. While \citet{Zhou2020} concentrated on a dipped arcade structure, solving only for the motions in the curved 2D, but static geometry of the magnetic field, we will here adopt an arch-shaped geometry and study the dynamics in a vertical plane where the field can still adjust dynamically.

In order to simulate the coronal rain formation and dynamics closer to the real situation, additional source terms for optically thin radiative losses and field-aligned thermal conduction are included in the governing equations in the existing MHD simulations. These terms need to be carefully treated, especially when the temperature gradient is steep. \citet{Bradshaw2013} pointed out that when encountering a steep temperature gradient, like in TR, simulations require a high resolution (typically 1 km or even 100 m), otherwise the evaporation might be underestimated, which will influence the temperature and density in the corona. In previous numerical simulations of prominence and coronal rain \citep[e.g.][]{Xia2012, Fang2013, Fang2015, Keppens2014, Xia2016, Xia2017}, the TR was not correctly resolved, as a consequence the evaporation cannot be triggered effectively. Several approaches have been proposed to correct the evaporation \citep{Lionello2009, Mikic2013, Johnston2017a, Johnston2017b}, and by combining ideas from previous works, \citet{Johnston2019} came up with the transition region adaptive conduction (TRAC) method for 1D hydrodynamic simulations, which is both efficient and accurate. \citet{Zhou2021} applied the TRAC method in multi-dimensional MHD simulations which allow dynamic AMR (Adaptive Mesh Refinement) and extended the TRAC idea with two new variants. 

In this paper, we present a 2.5D MHD simulation on coronal rain adopting turbulent heating patterns and the TRAC method, which displays the evolution and dynamics of coronal rain in more realistic situations. This paper is organized as follows. The governing equations and the numerical setup in our simulations are introduced in Sect.~\ref{sec:setup}. Sect.~\ref{sec:results} present the results of our simulations. Finally, the conclusions and discussion are in Sect.~\ref{sec:conc}.

\section{Numerical setup} \label{sec:setup}

Our simulation is performed using the parallelized Adaptive Mesh Refinement Versatile Advection Code \citep[MPI-AMRVAC,][]{Keppens2012, Porth2014, Xia2018, Keppens2021}. Disregarding the curvature of the solar surface, we use a cartesian ($x$, $y$) plane covering $-$ 30 Mm $\leq$ $x$ $\leq$ 30 Mm and 0 Mm $\leq$ $y$ $\leq$ 60 Mm, with gravity in the opposite direction of the $y$-axis. The effective resolution of our simulation which uses five AMR levels is 1536 $\times$ 1536, with the smallest grid size of 39 km in both directions.

\subsection{Governing Equations}

Our simulations are performed by solving MHD equations which include gravity, field-aligned heat conduction, 
parameterized heating terms and radiative cooling:

\begin{equation}
\frac{\partial{\rho}}{\partial{t}} + \nabla \cdot (\rho \bm{v})= 0,
\end{equation}
\begin{equation}
\frac{\partial{(\rho \bm{v})}}{\partial{t}} +\nabla \cdot  (\rho \bm{v} \bm{v} 
+ p_{tot} \bm{I} -  \frac{\bm{B}\bm{B}}{\mu_0} ) = \rho \bm{g},
\end{equation}
\begin{equation}
\frac{\partial{E}}{\partial{t}} +\nabla \cdot (E \bm{v} + p_{tot} \bm{v} - \frac{\bm{v} \cdot \bm{B}}{\mu_0} \bm{B}) 
= \rho \bm{g} \cdot \bm{v} + \nabla \cdot (\bm{\kappa} \cdot \nabla T) - R + H,
\end{equation}
\begin{equation}
\frac{\partial{\bm{B}}}{\partial{t}} + \nabla \cdot (\bm{v}\bm{B}-\bm{B}\bm{v})= 0,
\end{equation}
where $\bm{v}$, $\rho$, $\bm{I}$ and $\bm{B}$ are the velocity, plasma density, unit tensor and magnetic field, respectively. 
Assuming that plasma is completely ionized and the abundance of hydrogen and helium is 10:1, then we can obtain density $\rho = 1.4 m_p n_H$ with 
the number density of hydrogen $n_H$ and the proton mass $m_p$. The total pressure $p_{tot} = p + B^2 /2 \mu_0$ consists of thermal pressure $p$ and magnetic pressure, where $p$ is obtained by the ideal gas law $p = 2.3 n_H k_B T$. The gravitational acceleration is usually calculated by $\bm{g} = - g_{\odot} r_{\odot}^2 /(r_{\odot} + y)^2 \hat{y}$ where $r_{\odot}$ is solar radius, and $g_{\odot}$ = 274 m s$^{-2}$ is the solar surface gravitational acceleration. 
The $E = p/(\gamma - 1) + \rho v^2/2 + B^2 /2 \mu_0$ is total energy density with the ratio of specific heats $\gamma$ = 5/3. 
The anisotropic thermal conduction along the magnetic field lines is quantified by the term which contains $\bm{\kappa} = \kappa_{\parallel} \bm{bb}$, where the Spitzer conductivity $\kappa_{\parallel}$ equals 10$^{-6}$ $T^{5/2}$ erg cm$^{-1}$ s$^{-1}$ K$^{-1}$ and $\bm{b} = \bm{B} / B$ is the unit vector along the magnetic field. The radiative cooling term $R = 1.2 n^2_H \Lambda(T)$ is composed of the number density of hydrogen and the radiative loss function for optically thin emission $\Lambda(T)$, which is interpolated from the data given in \citet{Colgan2008} as demonstrated in previous simulations \citep{Xia2012, Xia2014, Keppens2014, Fang2015}. $\Lambda(T)$ is set to be zero below 10,000 K because the plasma there is no longer fully ionized and becomes optically thick. 
Many optically thin cooling curves have been implemented in MPI-AMRVAC, and \citet{Joris2021} studied the influence of these cooling curves on condensation formation and found that for different cooling curves, condensations may have different formation times and morphology. $H$ is the heating term which includes the background heating and localized heating $H = H_{bgr} + H_{loc}$. The background heating can resemble the photospheric source for coronal heating and balance the energy losses to help the corona reach a steady state. Here we adopt a vertical exponentially decaying background heating rate as follows: 
\begin{equation}
H_{bgr} = c_0 {\rm exp} \left ( - \frac {y}{\lambda_0} \right ),
\end{equation}
where $c_0$ = 10$^{-4}$ erg cm$^{-3}$ s$^{-1}$ and $\lambda_0$ = 50 Mm. 

For the localized heating $H_{loc}$, it is added after the system reaches a quasi-equilibrium. Here we adopted a randomized heating term to imitate the energy input from the lower solar atmosphere which has turbulent convection and moving magnetic features, similar to recent previous work \citep{Zhou2020}. The heating function is randomly distributed both in space and time, which can be expressed as $H_{loc} = c_1H(x)H(y)H(t)$, where $c_1$ equals 10$^{-2}$ erg cm$^{-3}$ s$^{-1}$. The randomized spatial distribution is imposed on the $x$ direction only, while $H(y)$ is set to be 
\begin{equation}
H(y)= 
\begin{cases}
{\rm exp}(- (y - y_c)^2/\lambda_1)  \qquad y > y_c\\
1  \qquad y \leq y_c
\end{cases}
\end{equation}
where $\lambda_1$ = 2 Mm$^2$ and $y_c$ = 3 Mm, representing the height of the TR in the final relaxed quasi-equilibrium system. The heating term in $x$ direction is a superposition of a series of waves of different wavelengths from the grid size to the simulation box: $H(x)= \Sigma_i A_i \sin(2 \pi x / \lambda_i + \phi_i))^2$, where $A_i$ is the amplitude, $\phi_i$ is the phase angle and the wavelength is defined by $\lambda_i = i(\lambda_{max} - \lambda_{min})/10^3 + \lambda_{min}$ where $\lambda_{min}$ is the grid size, $\lambda_{max}$ is twice the width of the domain and $i$ = 1, 2, 3, ..., 10$^3$. The heating in the solar atmosphere is considered to have a power-law spectrum, as a consequence we adopt $A_i \propto \lambda^s$, where the spectral index $s$ is chosen to be 5/6 \citep{Matsumoto2010}. \citet{Zhou2020} also mentioned that the choice of the spectral index didn't have too much impact on the results when 0.2 $\leq s \leq$ 2. Observational works provide evidence that heating in the solar corona is highly episodic \citep{Testa2014, Testa2020, Reale2019}. \citet{Testa2020} observed that two different peaks in 94 Å and 131 Å channels appear with 5 minutes intervals, which mark the presence of two distinct heating episodes. Large-scale corona simulations also reveal that chromosphere and hot corona could be maintained self-consistently when the convection zone is included in the simulation domain, and heating shows timescales from 2 to 5 minutes \citep{Hansteen2015}. Here the temporal distribution of the heating profile is an episodic Gaussian profile that has 5 min $\pm$ 75 s durations, attributed to the typical timescale and correlation time of the lower solar atmospheric motions. The localized heating term can then be expressed as the sum of a series of Gaussian functions with respect to time under these assumptions. To normalize the equations, we adopt normalization units of length $L_0$ = 10 Mm, temperature $T_0 = 10^6$ K and number density $n_0 = 10^9$ cm$^{-3}$. The other normalization units can be deduced accordingly. 

\subsection{Initial conditions and boundary conditions}
Following \citet{Fang2013, Fang2015}, we adopt a linear force-free magnetic configuration where all field lines make a constant angle of $\theta_0 = 30^\circ$ with the neutral line ($x$ = 0, $y$ = 0) for the initial magnetic field:
\begin{equation}
\begin{aligned}
B_x &= - B_0 \cos \left (\frac {\pi x}{L} \right ) \sin \theta_0 {\rm exp} \left (- \frac {\pi y \sin \theta_0}{L}\right ),  \\
B_y &= B_0 \sin \left (\frac {\pi x}{L} \right ) {\rm exp} \left (- \frac {\pi y \sin \theta_0}{L}\right ),   \\
B_z &= - B_0 \cos \left (\frac {\pi x}{L} \right ) \cos \theta_0 {\rm exp} \left (- \frac {\pi y \sin \theta_0}{L}\right ).
\end{aligned}
\end{equation}
Here $B_0$ = 20 G, and $L$ = 60 Mm is the horizontal size of our domain from $-$30 to 30 Mm.

In the initial state, the thermal structure below the transition region height of $h_{tr}$ = 2 Mm is a vertical distribution of temperature of about $T_{bo}$ = 8000 K at the bottom with smooth connection to $T_{top}$ = 1.5 $\times$ 10$^6$ K at the top using a hyperbolic tangent function:
\begin{equation}
T (y) = T_{bo} + 0.5 (T_{top} - T_{bo}) ( {\rm tanh}((y - h_{tr} - a)/w_{tr})+1) \qquad y \leq h_{tr},
\end{equation}
where $w_{tr}$ = 0.2 Mm and $a$ = 0.027. 
The temperature profile above the transition region height is given by
\begin{equation}
T (y) = [3.5 F_c (y - h_{tr}) / \kappa_{\parallel} + T^{7/2}_{tr}]^{2/7} \qquad y > h_{tr},
\end{equation}
where the vertical thermal conduction flux is $F_c$ = 2 $\times$ 10$^5$ erg cm$^{-2}$ s$^{-1}$, and $T_{tr}$ =1.6 $\times$ 10$^5$ K \citep{Zhou2018}. 
The number density at the bottom is set to be 1.151 $\times$ 10$^{15}$ cm$^{-3}$, then the initial density is derived under the assumption of hydrostatic equilibrium.
The initial velocity is set to be zero.

For boundary conditions, two grid layers exterior to the domain are used as ghost cells. We use a fixed zero velocity, fixed magnetic field, and fixed gravity stratification of density and pressure predetermined in the initial condition for the bottom boundary. At the left and right physical boundaries, density, energy, $y$ component of momentum, $B_y$ are set symmetrically, while $x$ and $z$ components of momentum, $B_x$ and $B_z$ are adopted asymmetrically to ensure zero face values. As for the top conditions, we use a fixed zero velocity and employ a separate pressure-density extrapolation from gravity stratification as introduced in the initial condition. We determine $B_x$ and $B_z$ using a zero-gradient magnetic field extrapolation and derive $B_y$ with a second-order centered difference evaluation of zero-divergence constraint. 
 
 \subsection{Numerical methods}

We use MPI-AMRVAC\footnote{http://amrvac.org} to solve the governing equations adopting a three-step Runge-Kutta time integration with a third-order slope limited reconstruction \citep{Cada2009} and Harten-Lax-van Leer Riemann solver. To fulfill that the magnetic field divergence is close to zero, we employ the upwind constrained transport (CT) method which uses staggered grids for the magnetic field, which was applied in the companion general relativistic MHD code variant BHAC \citep{Gardiner2005, Porth2017}. Radiative cooling and field-aligned thermal conduction are handled with the exact integration method \citep{Townsend2009} and the Runge-Kutta Legendre super-time-stepping (RKL-STS) scheme \citep{Meyer2012, Meyer2014}, respectively. As mentioned in Sect.~\ref{sec:intro}, we adopt the TRAC method to correct the evaporation. The basic idea is defining a cutoff temperature $T_c$, and then modifying the local evaluations of radiative cooling and thermal conduction terms at the places which have a temperature lower than $T_c$ \citep{Lionello2009, Mikic2013, Johnston2017a, Johnston2017b, Johnston2020, Johnston2019}. In this paper, we employ the basic 1D TRAC method, and the local cutoff temperature inside each block is calculated separately when it is used for 2D or 3D simulations. It is worth mentioning that a field line-based TRACL method and a block-based TRACB method \citep{Zhou2021}, as well as the LTRAC method proposed by \citet{Iijima2021}, can now also be used for multi-D simulations inside MPI-AMRVAC.

\begin{figure}
\includegraphics[trim = 0mm 0mm 0mm 0mm, clip, width=1.0\textwidth]{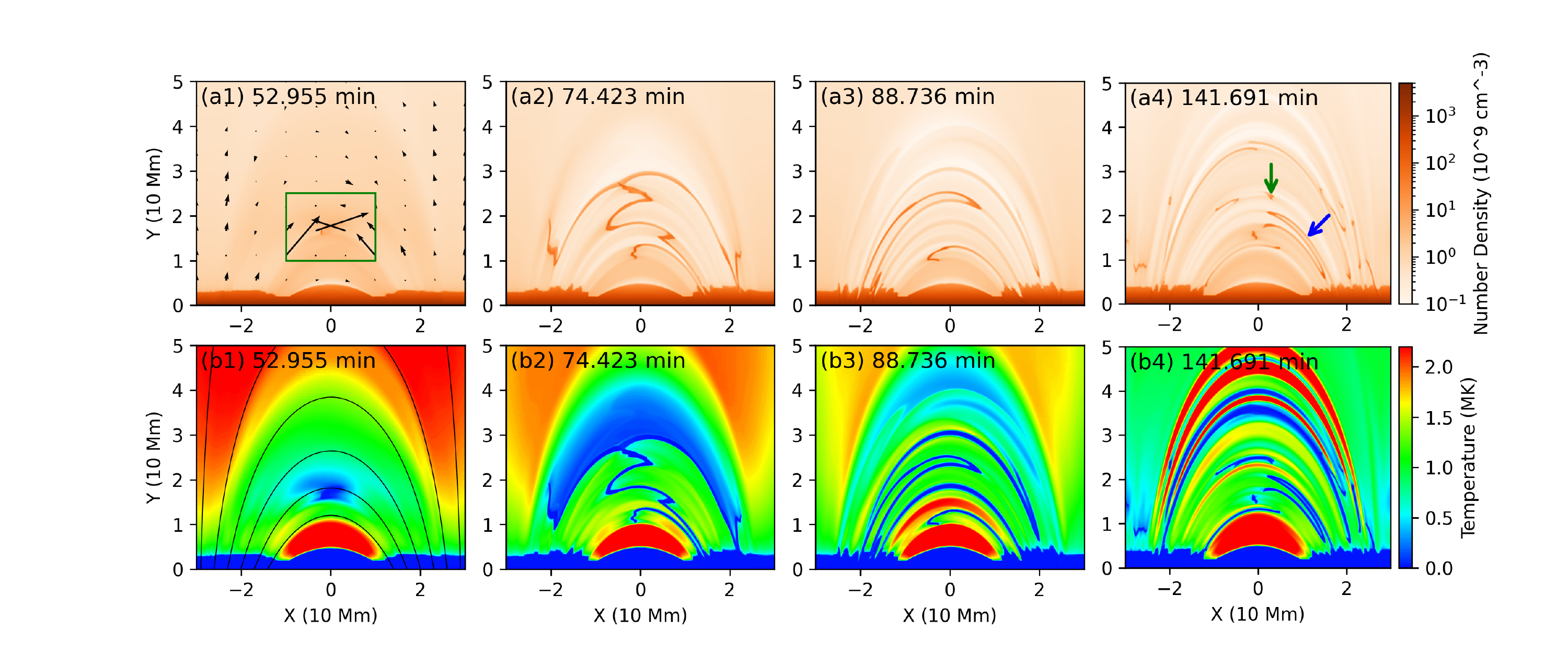}
\caption{Snapshots at different time showing the evolution of the condensation. The upper and lower panels display the number density and temperature respectively. The black arrows in panel (a1) denote the velocity, and the green rectangle outlines the field-of-view of Figure~\ref{fig2}. In panel (a4), the green and blue arrows indicate the small and large condensations, respectively. The black curves in panel (b1) represent the magnetic field lines. 
An animation (Animation1.mp4) of this figure is available, showing the evolution of number density and temperature from $t$ = 0 min to $t$ = 662.655 min. The time cadence of the animation is about 1.431 min.
\label{fig1}}
\end{figure}

\section{Results} \label{sec:results}

\subsection{Overall evolution, typical rain blob formation and descend} \label{sec31}

\begin{figure}
\includegraphics[trim = 0mm 0mm 0mm 0mm, clip, width=1.0\textwidth]{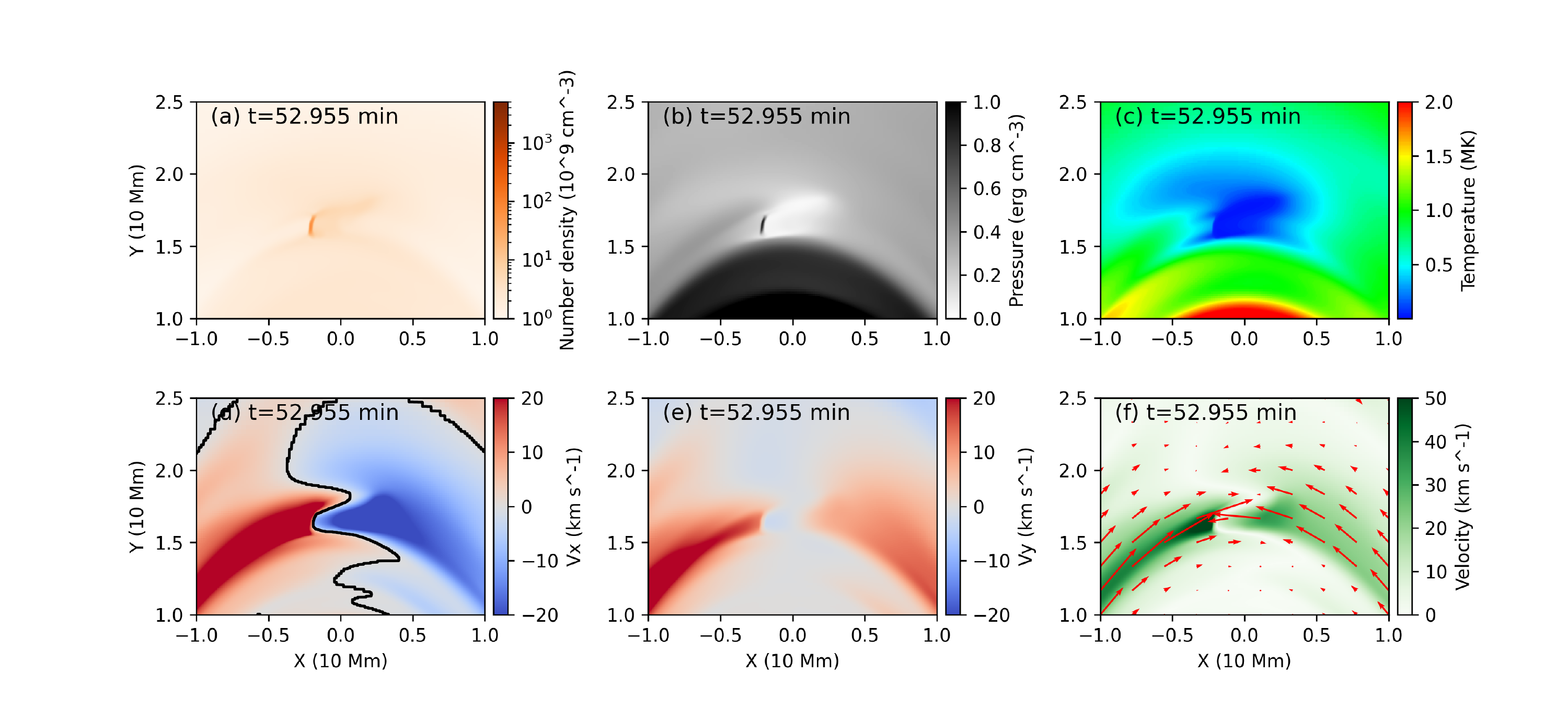}
\caption{The number density, pressure, temperature and velocity plot of the first blob forming in a zoomed view denoted by the green rectangle in Figure~\ref{fig1}(a). The black contour in panel (d) denotes the position where $V_x$ equals 0. 
\label{fig2}}
\end{figure}

To help the system with the above initial setup reach a quasi-equilibrium state, we run the simulation with $H = H_{bgr}$ for 171.6 min and make sure that the maximal residual velocity is below 5 km s$^{-1}$. Beginning with this equilibrated system, we reset the time to zero and include the localized heating $H = H_{bgr} + H_{loc}$ to the simulation. The localized heating evaporates plasma and causes upflows into the corona, and the densities around the loop apexes continue to increase. Meanwhile, the temperature first increases but then starts to reduce slowly (see Animation1.mp4). This reduction happens as radiative cooling starts to dominate over heating and thermal conduction, which leads to the onset of thermal instability. After about 53 minutes of localized heating, near the loop top region, a small condensation segment with temperature of around 0.02 MK and number density of 6.8 $\times$ 10$^{10}$ cm$^{-3}$ suddenly forms, which is shown in the left panels in Figure~\ref{fig1}. Figure~\ref{fig2} exhibits the details of a loop top region denoted by the green rectangle in Figure~\ref{fig1}(a1). In this loop top region, temperature and gas pressure decline significantly, causing matter drawn in from the surroundings. The induced siphon flows coming from two sides with speeds up to 57 km s$^{-1}$ form the condensation segment, which has a length of 1.5 Mm (see Figure~\ref{fig2}). Similar condensation processes continuously take place and extend into strands on both sides of the first affected coronal loop, which was ascribed to a common footpoint heating process \citep{Antolin2012, Fang2013, Fang2015}. The large-scale condensations therefore exhibit a zigzag shape as shown in panels (a2) and (b2) of Figure~\ref{fig1}. As the condensations continue to develop, they lose their balance spontaneously and start to descend slowly along either side of the magnetic field lines and split into several smaller blobs. After this large-scale condensation process, condensations continuously occur at nearly every place of the coronal loops, which seems more disorganized and less vigorous (see panel (a4)). 

\begin{figure}
\includegraphics[trim = 0mm 20mm 0mm 20mm, clip, width=1.0\textwidth]{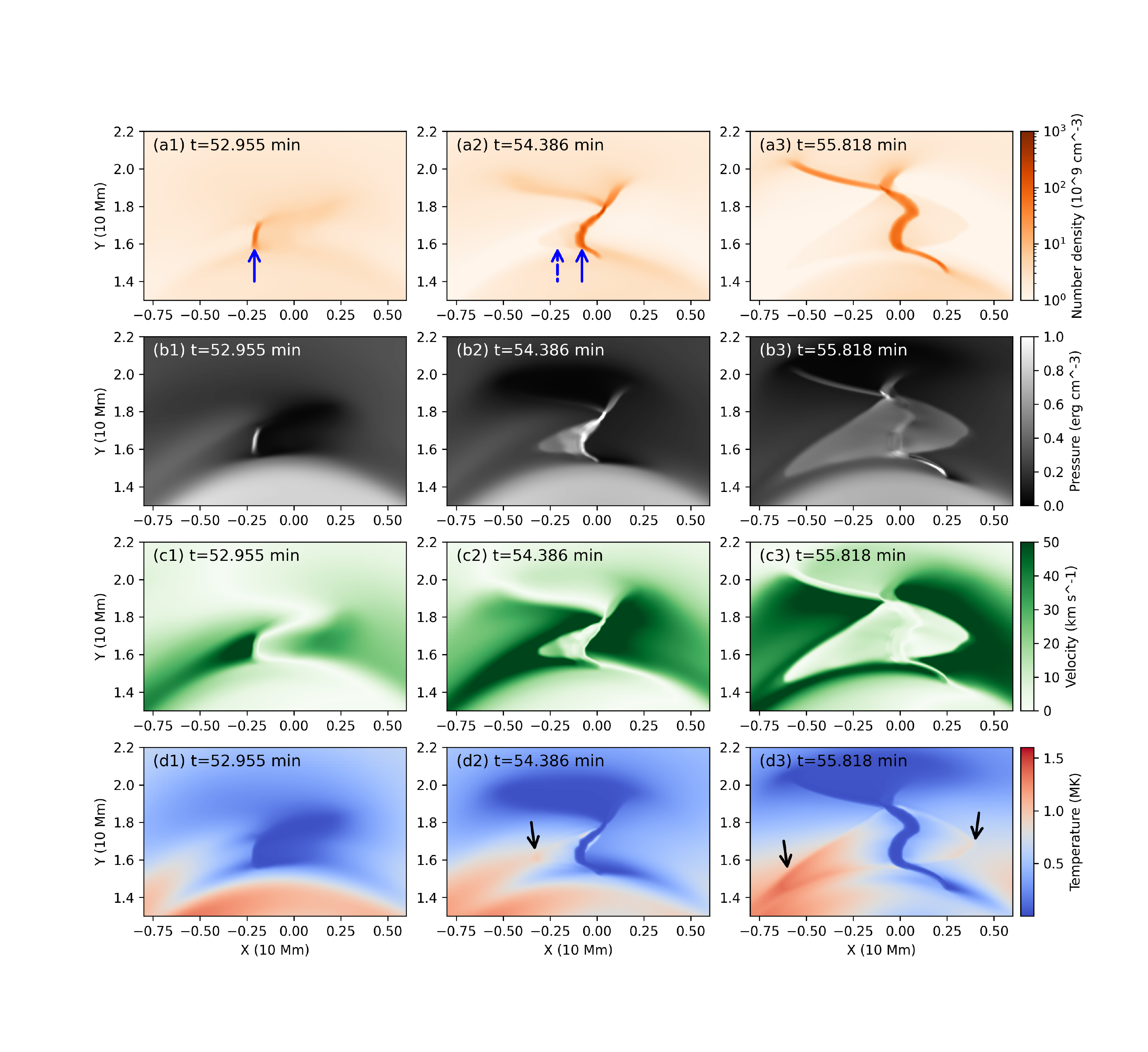}
\caption{The number density, pressure, plasma velocity magnitude, and temperature maps 
showing the evolution of two-sided rebound shocks. The solid blue arrows in panels (a1) and (a2) 
indicate the present position of the condensation while the dashed arrow in panel (a2) denotes 
the position at last snapshot copied from panel (a1). In panels (d2) and (d3), the black arrows
indicate the heating near the shock fronts.
\label{fig3}}
\end{figure}

During the first condensation, we observe similar phenomena such as siphon flows and rebound shocks compared to earlier simulations in \citet{Xia2011, Xia2012, Xia2017} and \citet{Fang2013, Fang2015}, as presented by the temporal evolution of number density, gas pressure, plasma velocity magnitude, and temperature maps in Figure~\ref{fig3}. Panels (a1) and (b1) show the dense plasma blob formed by the siphon flows, and its pressure is much higher than the surroundings. The pressure at the right side of the blob is much lower than the left side. The velocity magnitude map in panel (c1) reveals that the siphon flow at the left of this condensation has a higher speed of 67 km s$^{-1}$, compared to the right siphon flow at a speed of around 10 km s$^{-1}$. From 52.955 min to 54.386 min, the location of this condensation moved to the right for about 1.0 Mm in the presence of pressure gradient and stronger siphon flow at its left, as shown by the blue arrows in panel (a2). The solid arrow indicates the present position of the condensation while the dashed one denotes the position at the previously shown snapshot. Meanwhile, one left rebound shock was generated and propagated against the continuous siphon flows along the magnetic field. The shock front displays a fan-shaped structure, as the processes of condensation and shock formation on adjacent magnetic field lines start at different times, which first happen at the center and then propagate outward from the blob. At this instant, the shock compresses the plasma from 1.8 $\times$ 10$^9$ to 3 $\times$ 10$^9$ cm$^{-3}$, decreases the inflow speed from about 67 km s$^{-1}$ to around 20 km s$^{-1}$, and heats the plasma from 0.1 MK to 0.6 MK (see panels (a2)$-$(d2)). About 1.43 minutes later, the right-directed rebound shock is also generated and begins to propagate and heat the plasma (see panels (a3)$-$(d3)). As the left rebound shock occurs earlier, its front moves further than the right one. The plasma near both fronts becomes hotter, as denoted by the black arrows in panel (d3). 

\begin{figure}
\centering
\includegraphics[trim = 0mm 0mm 0mm 0mm, clip, width=0.9\textwidth]{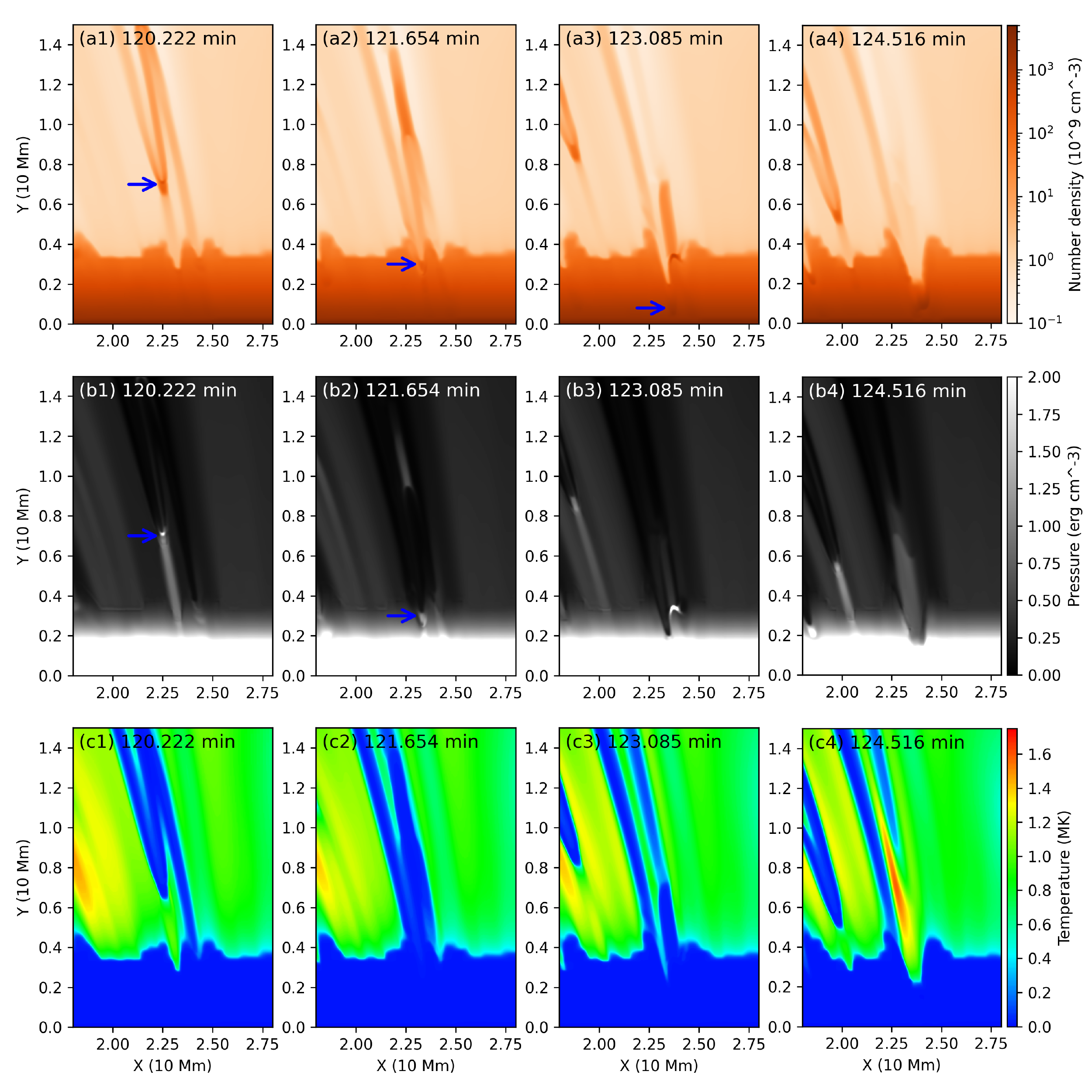}
\caption{The number density, pressure and temperature maps showing the chromosphere and transition region variations when a blob falls down and enters the lower solar atmosphere. The blue arrows point out the positions of the blob we focus on.
\label{fig4}}
\end{figure}

As the blobs lose their balance and move downwards, they eventually hit the TR and enter the lower chromosphere, and we find that nearly all of them trigger concurrent upflows which follow the same loops as the downflows. Panels (a1)$-$(a4) in Figure~\ref{fig4} show the number density map between 120.222 min and 124.516 min when a falling blob enters the TR, while panels (b1)$-$(b4) and (c1)$-$(c4) display the corresponding pressure and temperature change. The blue arrows point out the positions of the blob we focus on. As the blob falls downwards, it compresses the plasma on its way, which makes the pressure in front of it increase, and the pressure in its wake is greatly reduced. After the blob reaches the TR, the local density and pressure increase, which leads to an upward force, and plasma is sent back to the corona. These upflows move along the same coronal loops as the coronal rain blobs, and the plasma in the loops is heated to a higher temperature of 1.51 MK, as shown in panel (c4). The similar phenomenon has been observed in observations and simulations \citep{Tripathi2009, Antolin2010, Kleint2014, Fang2015}. \citet{Fang2015} found that before the rebound shock and upflows reach far into the loop, the temperature of the low-density loop increases due to the background heating. Then the rebound shocks and upflows produced by the impact of the rain heat the low-density loops efficiently to a high temperature, and the heating follows the movement of the upflows. Further upflows coming from evaporation due to the footpoint heating will also heat the loop afterward, which can be seen from the attached animation (Animation1.mp4). 

\begin{figure}
\centering
\includegraphics[trim = 0mm 10mm 0mm 0mm, clip, width=0.8\textwidth]{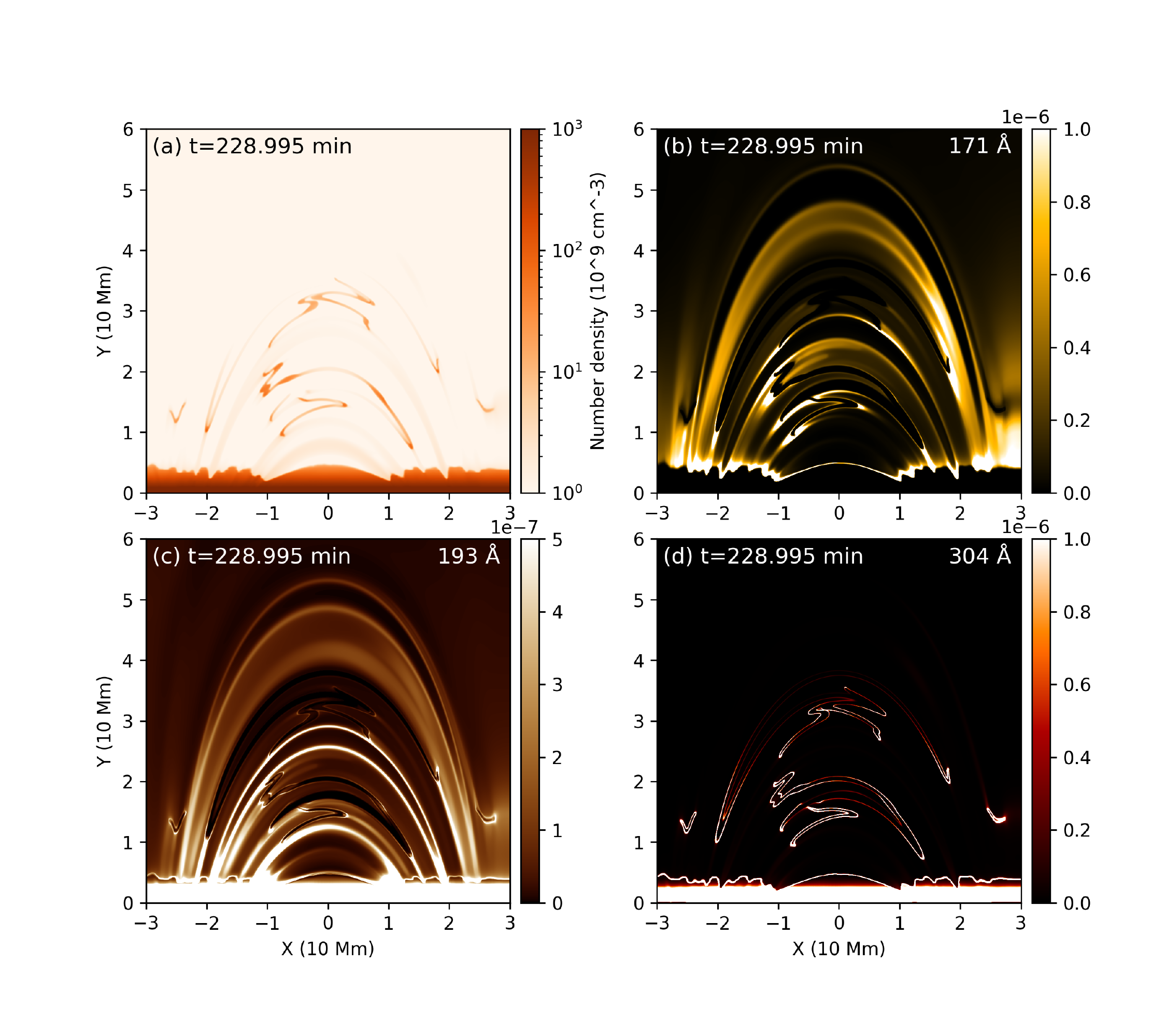}
\caption{The number density, synthesized EUV 171 {\AA}, 193 {\AA} and 304 {\AA} images showing the coronal rain blobs. 
\label{fig5}}
\end{figure}

After the first condensation process, the whole system gets into a constant cycle where both small and large blobs coexist, as displayed in Figure~\ref{fig1}(a4). The green and blue arrows point to a small and large condensation, respectively. Large clumps consisted of numerous condensations are usually denoted as ``coronal rain showers" \citep{Antolin2012}. To compare the results with observed images from the Atmospheric Imaging Assembly (AIA) instrument onboard of Solar Dynamics Observatory, Figure~\ref{fig5} displays the number density and synthetic EUV images of the coronal loops and coronal rain at 228.995 min in our simulation. The technique to obtain synthetic images like this is described in \citet{Xia2014}. Here we choose the 171 {\AA}, 193 {\AA}, and 304 {\AA} wave channels, which have main contribution temperatures of around 0.8, 1.5, and 0.08 MK, respectively. In the number density map (see panel (a)), we can clearly see that condensed plasma is connected to others and combined into large clumps. The hot 171 {\AA} and 193 {\AA} channel images in panels (b) and (c) distinctly show the bright coronal loops and the dark coronal rain clumps embedded in them. In the 304 {\AA} channel, the coronal rain clumps look bright at the boundary and dark inside, implying that these clumps have a core with a temperature around or even lower than 10$^4$ K, which is consistent with the observations of coronal rain showers \citep{Antolin2015}.

\subsection{Statistical analysis} \label{sec32}

\begin{figure}
\centering
\includegraphics[trim = 0mm 0mm 0mm 0mm, clip, width=0.9\textwidth]{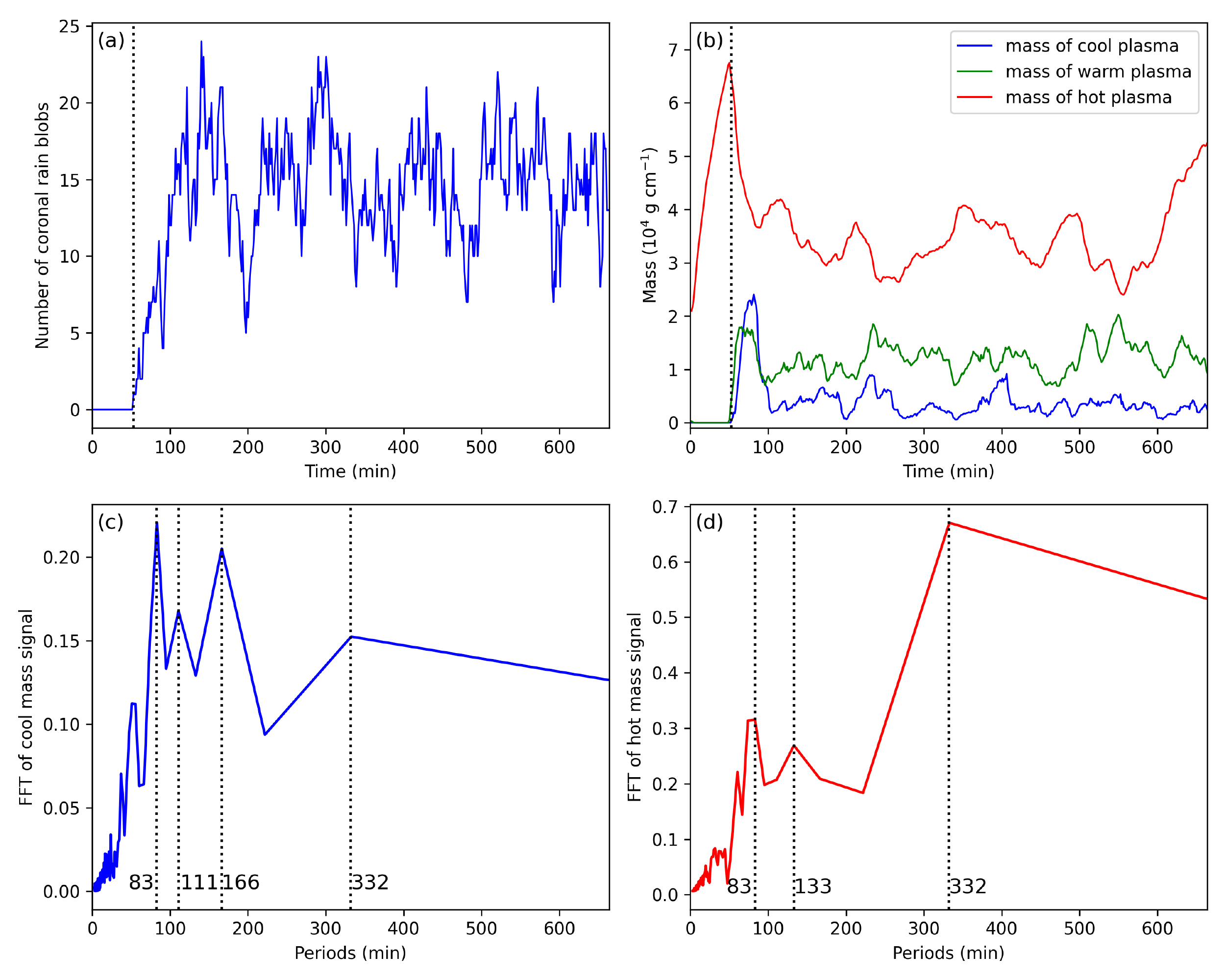}
\caption{Time evolution of (a) the number of all blobs and (b) the mass of cool, warm and hot plasma in the corona. Panel (c) shows the FFT result of the mass of cool plasma, and panel (d) displays the FFT result of the mass of hot plasma.
\label{fig6}}
\end{figure}

To explore the dynamics of the coronal rain blobs, we count all the blobs which appeared during the simulation and study their properties statistically. We use the former standard \citep{Hirayama1985, Muller2005, Antolin2010} to define a grid cell located above the chromosphere-corona-transition-region as a component of the coronal rain blobs if its number density is larger than 7 $\times$ 10$^9$ cm$^{-3}$ and has a temperature beneath 1 $\times$ 10$^5$ K. Regarding all connected cells as one coronal rain blob, we identify the coronal rain blobs and collect their characteristics, such as number, mass, width, length, area, centroid location, and centroid velocity at each saved snapshot. Using this identification method, we find that the total number of blobs is 6060 in all 427 snapshots (the number of snapshots for which we chose to save all data at a cadence of $\Delta t= 1.431$ min from the condensation start till the end of simulation), which does of course count blobs that moved location multiple times. There is a maximum of 24 blobs in one snapshot. The temporal change of the number of coronal rain blobs at any instant is shown in Figure~\ref{fig6}(a). Panel (b) displays the variation of the mass of hot, warm, and cool plasma in the corona over time. The ``cool" plasma refers to the plasma with a temperature below 0.02 MK, and the ``warm" plasma is the plasma between 0.02 MK and 0.5 MK, and the plasma hotter than 0.5 MK is identified as ``hot" plasma. As soon as the simulation starts, the mass of hot plasma begins to increase continuously, and it increases by about 4.5 $\times$ 10$^4$ g cm$^{-1}$, almost tripled within 53 minutes. At around 53 min, as the mass of hot plasma decreases, the mass of warm plasma begins to increase, and then the mass of cool plasma also increases, which illustrates that catastrophic cooling processes cool down hot plasma in situ in the corona, and the first condensation formed. From $t$ $\sim$ 75 min, blobs begin to enter the TR, the mass of cool plasma reaches its first peak and then starts to decrease. After blobs hit the chromosphere, there are upflows entering the relatively empty loops, heating the loop as presented in Figure~\ref{fig4}, so the mass of hot plasma has a tendency to increase. Then the number and mass curves both seem to have several cycles, which has been investigated in previous observational works \citep{Antolin2012, Kohutova2016, Froment2020} and simulations \citep{Muller2003, Fang2015}. Here we don't see any time gap between these cycles, i.e., the coronal rain blobs continue to occur all the time, different from the results from \citet{Fang2015}. This is likely a direct result of the randomized heating pattern we adopted, as a result, the chromospheric evaporation rate is not constant and has a different spatial distribution. This difference indeed confirms earlier studies on the basis of observations, where one speculated that the temporal behavior of coronal rain cycles may well reflect the heating prescription at work, and thus may provide clues to the coronal heating paradigm \citep{Antolin2010}. Although the continuous, but randomized heating imposed on our arcade footpoints has a spectral behavior with a peak of about 5 minutes, the time sequence displayed in Fig.~\ref{fig6} rather shows peaks at much longer timescales: Fast Fourier Transform (FFT) results of the cool plasma mass signal (blue curve) and the hot plasma mass signal (red curve) show main periodicities of 83 minutes (see panels (c) and (d)). Longer ones such as the peak at 332 minutes may be meaningless, as the FFT is not accurate enough due to the insufficient simulation time. 

\begin{figure}
\centering
\includegraphics[trim = 10mm 10mm 10mm 0mm, clip, width=1.0\textwidth]{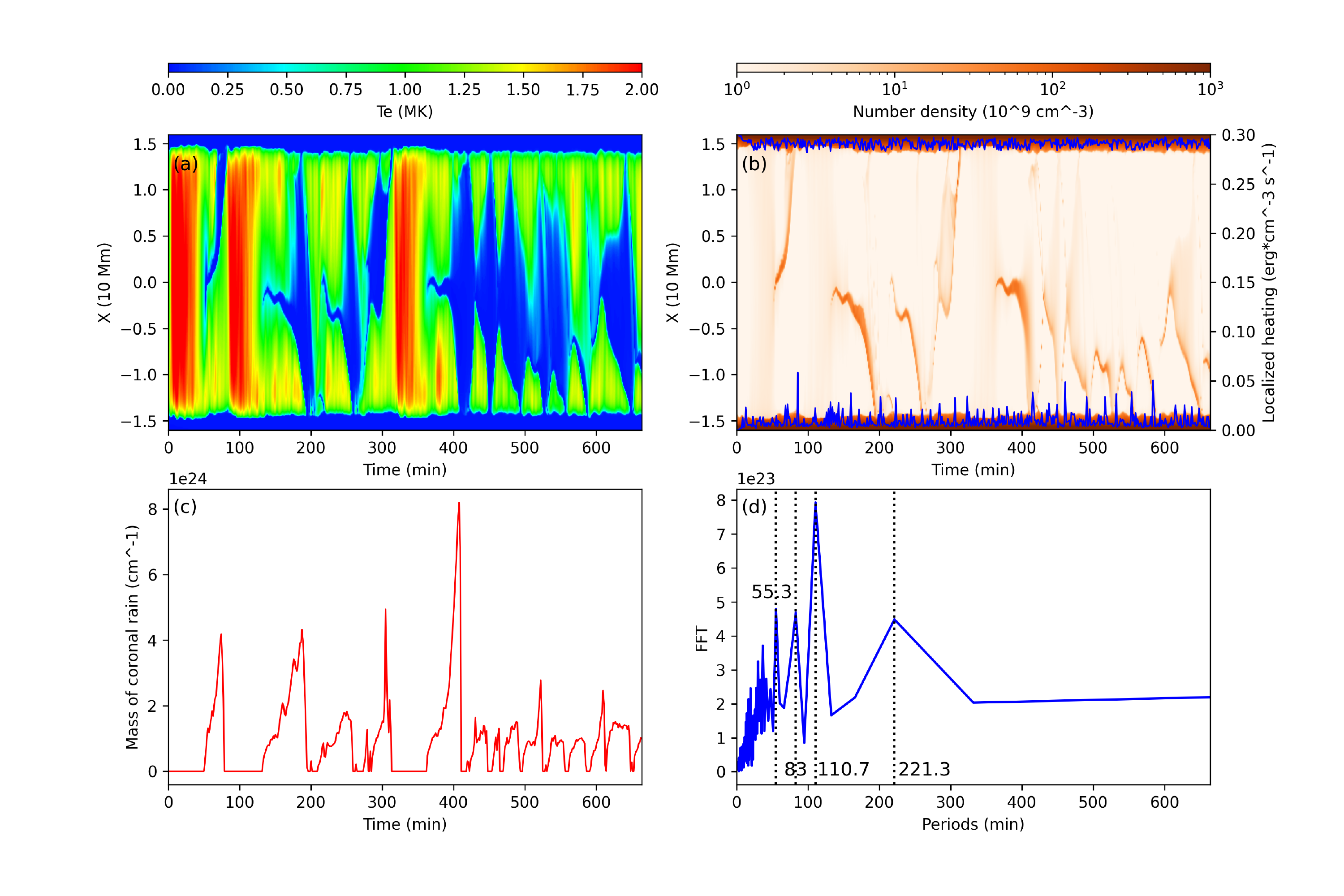}
\caption{The time evolution of temperature (panel (a)) and number density (panel (b)) in one magnetic field line whose left footpoint is located at $x$ = 16 Mm during the simulation. The localized heatings at two footpoints are presented by the blue curves in panel (b), with the lower curve denotes the original value at the left footpoint, and the upper curve is 0.3 minus the value at the right footpoint. Panel (c) displays the time evolution of mass of the coronal rain blobs in this magnetic field line, and the FFT result is shown in panel (d).
\label{fig7}}
\end{figure}

We track several different field lines during the whole simulation, and the temperature and number density evolution along the magnetic field line passing the footpoint at $x$ = 16 Mm are demonstrated in Figures~\ref{fig7}(a) and~\ref{fig7}(b), respectively. The blue curves in panel (b) represent the change of localized heating at two footpoints, with the lower curve exhibiting the original value at the left footpoint, and the upper curve is drawn by 0.3 minus the value at the right footpoint. We can see that as soon as the localized heating starts, the loop is heated to a temperature of several MK. Around tens of minutes later, the loop begins to cool down, and the temperature drops to transition region and chromospheric temperature in approximately 20 minutes, meanwhile the loop density gradually increases. Suddenly the condensations form, move inside the loop, and eventually enter the lower atmosphere. Then the loop continuously undergoes heating and cooling, and condensations occur from time to time. Panel (c) displays the time evolution of the mass of coronal rain blobs on this magnetic field line, from which we can obtain that the coronal rain in a single field line experiences several cycles. The FFT result in panel (d) illustrates that the coronal rain mass on this field line has several period components, similar to the period of the whole system. Comparing the results of different field lines, we find that different field lines have similar periods, which implies that a synchronizing mechanism is acting across the field lines, making coronal rain occur near-simultaneously across neighboring field lines rather than purely driven by the field line's own heating, as suggested by \citet{Antolin2012} and \citet{Antolin2020}. This synchronizing mechanism could be purely due to the magnetic and gas pressure variations across the field lines, as shown by \citet{Fang2013, Fang2015}. A single loop or adjacent magnetic loops experience typical heating-condensation cycles repeatedly: heating leads to chromospheric evaporation and then radiative cooling accompanied by thermal instability and condensation of the plasma. For the whole system, it undergoes a continuous heating-condensation cycle, alternating heating, radiative cooling resulting in the subsequent plasma condensations, which is then followed by long competing phases during which the heating and cooling keep happening and the condensations occur all the time. Changes in plasma properties are communicated across field lines by fast magnetosonic waves. 

\begin{figure}
\includegraphics[trim = 40mm 10mm 30mm 20mm, clip, width=1.0\textwidth]{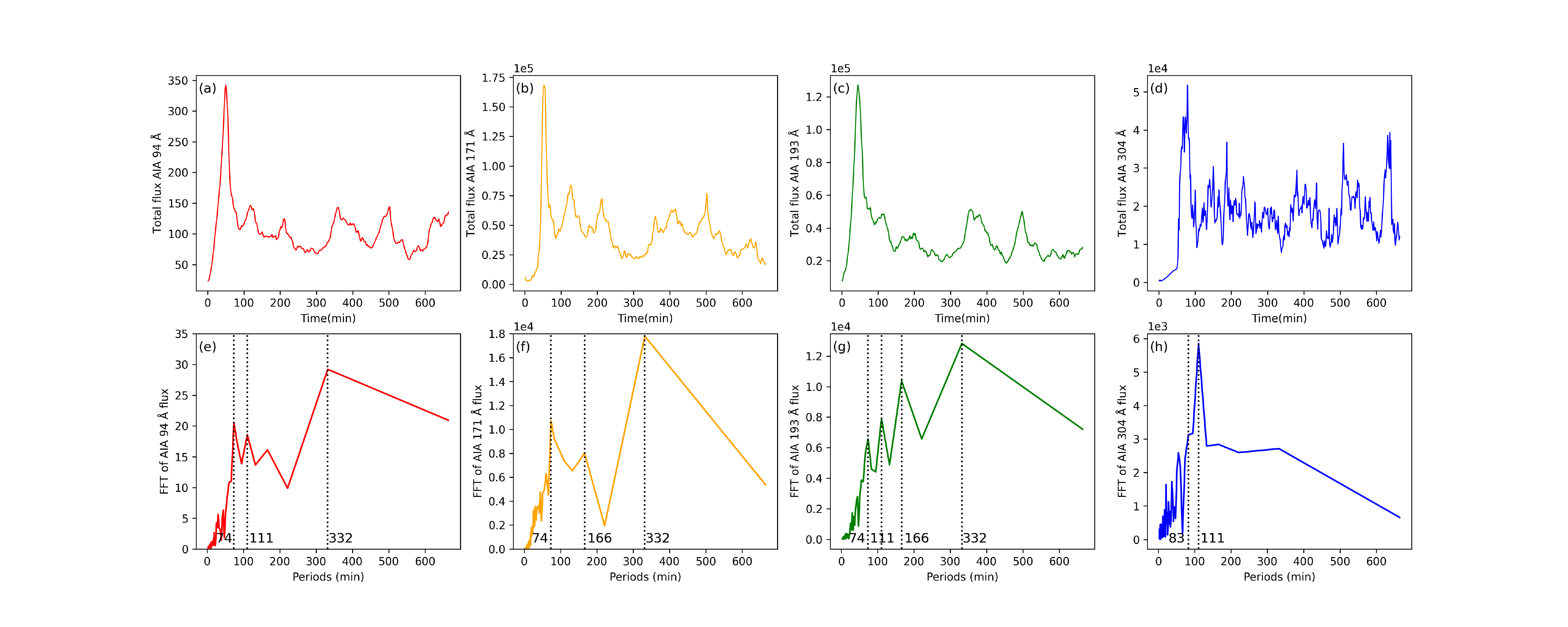}
\caption{Time evolution of EUV emission. Panels (a) $-$ (d) display the total flux of synthesized AIA 94 {\AA}, 171 {\AA}, 193 {\AA} and 304 {\AA} emissions, and panels (e) $-$ (h) show the FFT results of the total flux signals.
\label{fig8}}
\end{figure}

Recent observational studies reveal that EUV intensity pulsations with periods of several hours are a common phenomenon in solar coronal loops \citep{Auchere2014, Auchere2016, Froment2015}, and they are interpreted as signatures of evaporation $-$ condensation cycles \citep{Froment2017, Froment2018}. In order to compare our results with observations, we also calculate the intensity of the synthetic EUV images. The total flux of the AIA 94 {\AA},  171 {\AA}, 193 {\AA}, and 304 {\AA} emissions in the corona as time curves are presented in Figures~\ref{fig8} (a) $-$ (d). The AIA 94 {\AA} channel has strong responses to temperatures at about 6.8 MK, so it reflects the change of the mass of hot plasma. As mentioned before, the 193 {\AA} and 171 {\AA} channels also have the main contribution from hot and warm plasma. As a result, the AIA 94 {\AA} and 193 {\AA} flux increase immediately after the localized heating starts and have similar morphological changes with the mass curve of hot plasma in Figure~\ref{fig6} (b).  Compared to these hot channels, the start time of the 171 {\AA} emission is delayed. The intensity of AIA 304 {\AA} reflects the emissions of cold coronal rain, which increases drastically after tens of minutes. All these channels display cycles with similar periods compared to the mass of coronal rain (see panels (e) $-$ (h)).

\begin{figure}
\centering
\includegraphics[trim = 0mm 0mm 0mm 0mm, clip, width=0.95\textwidth]{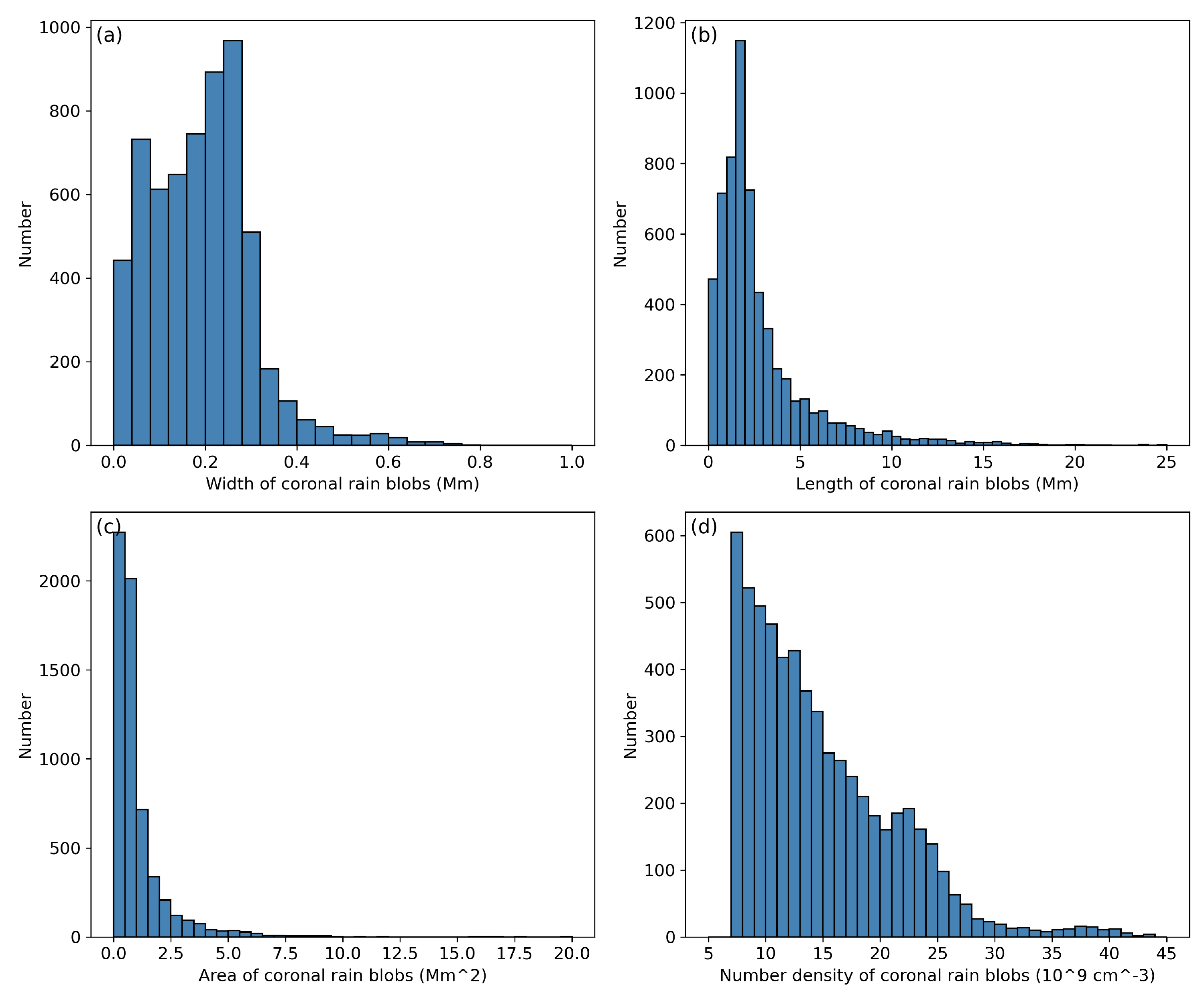}
\caption{Distribution of width, length, area and mean number density of coronal rain blobs.
\label{fig9}}
\end{figure}

Figure~\ref{fig9} displays the distributions of width, length, area, and mean number density of our simulated coronal rain blobs. The width of each blob is calculated from the diameter of its biggest inscribed circle. Most blobs have widths less than 800 km, as shown in panel (a). The length of each blob is calculated by half the perimeter of the blob minus its width, to mimic the curved length along the clump's spine. The lengths of blobs distribute broadly from a few hundred km to 25 Mm, with the bulk peaking at 1.5 $-$  2.0 Mm. As there are large condensations and irregularly shaped blobs, the area may describe the scale of the blob more accurately, as shown in panel (c). The areas of most blobs are under 5 Mm$^2$, and blobs are dominated by smaller ones with areas under 0.5 Mm$^2$. The mean number density of the blobs is ranging from 7 $\times$ 10$^9$ $-$ 4.4 $\times$ 10$^{10}$ cm$^{-3}$ (see panel (d)). The results are in good agreement with previous studies \citep{Fang2013, Antolin2015, Antolin2020}. Due to the limit of the simulation resolution, small-scale condensations may be not observable, which is considered as the tip-of-the-iceberg effect and therefore suggests a number increase for blobs with smaller sizes for higher resolution \citep{Antolin2012, Fang2013, Scullion2014}. We may study even more highly resolved rain dynamics in follow-up work, with the distinct possibility to identify small-scale, secondary fluid instabilities at work.

\begin{figure}
\includegraphics[trim = 0mm 0mm 0mm 0mm, clip, width=1.0\textwidth]{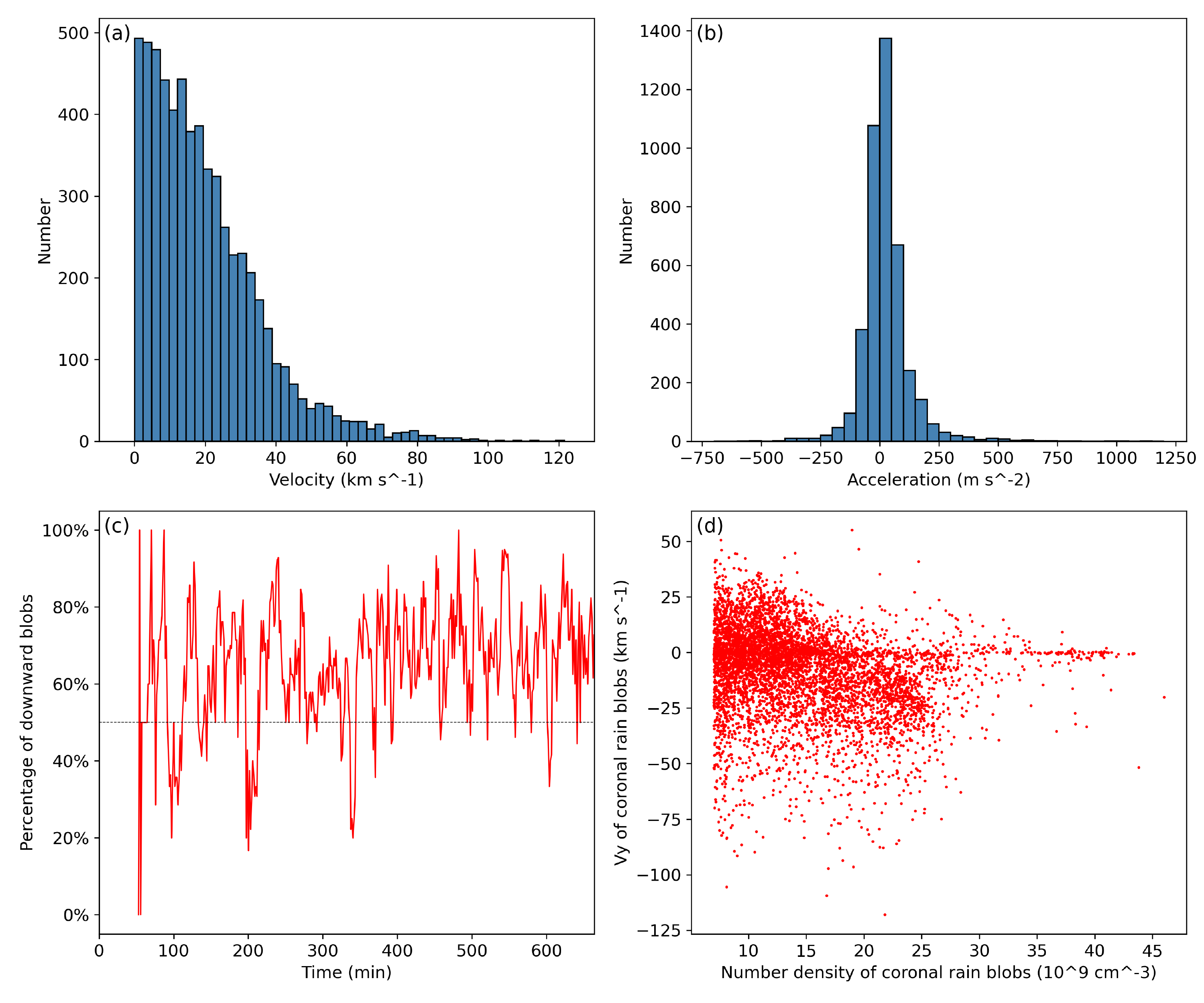}
\caption{Statistical results of coronal rain plasma properties on (a) distribution of velocity; (b) distribution of acceleration; (c) change of percentage of downward blobs over time and (d) scatter plot of the $y$-velocity components and the mean number densities of all the blobs.
\label{fig10}}
\end{figure}

Dynamic properties of coronal rain plasma are analyzed and shown in Figure~\ref{fig10}. The distributions of velocity and acceleration are displayed in panels (a) and (b). The velocities of the coronal rain blobs have a wide distribution varying from several km s$^{-1}$ to above 120 km s$^{-1}$, with a mean speed of 22.4 km s$^{-1}$. Most blobs have a speed under 40 km s$^{-1}$. The acceleration of the majority of the blobs is between $-$200 and 200 m s$^{-2}$, and is clustered around 50 m s$^{-2}$. The measured acceleration is typically below the gravitational free-fall acceleration, which has been pointed out and explained by pressure-mediated effects in \citet{Muller2003, Antolin2010} and \citet{Fang2015}. Panel (c) shows the percentage of downward blobs ($V_y <$ 0) as a function of time. We can see that although most blobs are falling downward, there are blobs moving upwards at basically every moment. Sometimes the number of rising blobs is even higher. The scatter plot in panel (d) displays the $y$-velocity components and the mean number densities of the blobs. We can see that as the number density increases, fewer blobs are moving upward, and their upward speeds also decrease. For the denser blobs, they hardly move upwards. This can be explained since a blob of larger density requires a larger pressure gradient to overcome the acceleration of gravity and move upwards, which has also been mentioned in previous studies \citep{Oliver2014, Martinez2020}.

\begin{figure}
\centering
\includegraphics[trim = 0mm 0mm 0mm 0mm, clip, width=0.6\textwidth]{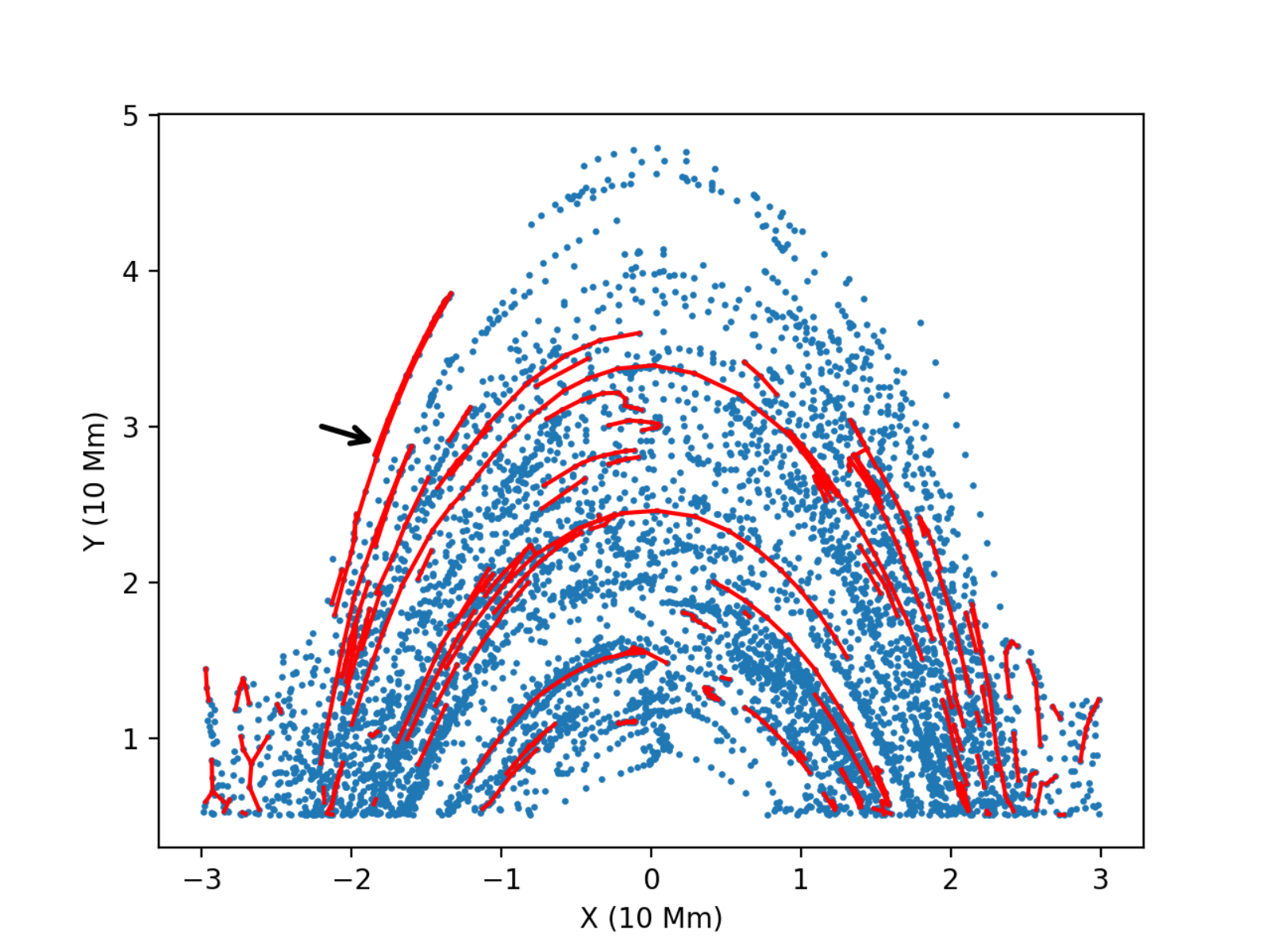}
\caption{Scatter plot showing the positions of all blobs by their horizontal and vertical centroid locations. Individual blue dots indicate blob positions through the entire simulation. Red curves connect those dots that are actually the same blob evolving in time, and thus display trajectories between 133.103 min and 194.646 min.
\label{fig11}}
\end{figure}

\subsection{Dynamics of individual blobs}

\begin{figure}
\centering
\includegraphics[trim = 0mm 15mm 0mm 30mm, clip, width=0.75\textwidth]{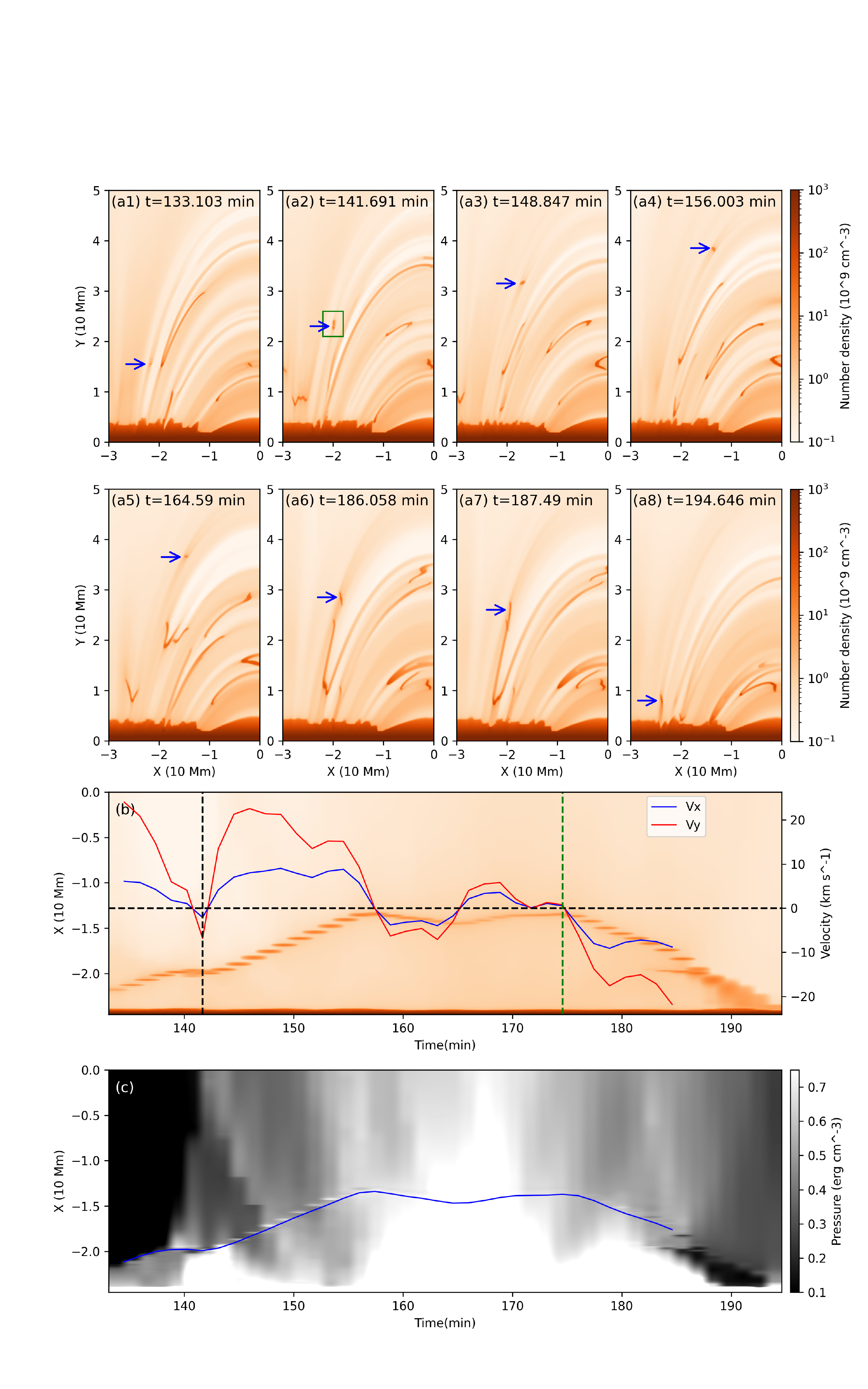}
\caption{Movement of a blob. Panel (a) is the number density map time series in which the blob we focus on is denoted by blue arrows. The green rectangle in panel (a2) outlines the FOV of Figure~\ref{fig13}. Panels (b) and (c) show the time evolution of the number density and pressure along the field line where the blob is located, respectively. In panel (b), the blue (red) line describes the change of the $x$-velocity ($y$-velocity) component of this blob. The black dashed vertical line and green dashed vertical line correspond to $t$ = 141.691 min (simultaneously with panel (a2) and Figure~\ref{fig13}) and $t$ = 174.609 min, respectively. In panel (c), the blue curve denotes the trajectory of this blob.
\label{fig12}}
\end{figure}

As we have the properties such as positions and velocities of all blobs in the simulation, we can track the movement of individual blobs. Here we use the position and velocity of a blob to predict the position where it may appear at the next moment, and look for the blob appearing near this position. The spatial distribution of all blobs in the simulation is displayed in Figure~\ref{fig11}. We can see that the blobs appear nearly everywhere in the magnetic loops. In order to make the figure more concise, only the trajectories of the blobs between 133.103 min and 194.646 min are indicated by the red curves. Most blobs fall down along the magnetic loops, while there are still many blobs moving upwards as quantified in Figure~\ref{fig10}(c), and some of them even pass the top of the loop to the other side. Figure~\ref{fig12}(a) is the number density map showing the movement of a blob in detail, denoted by the black arrow in Figure~\ref{fig11} and blue arrows in Figure~\ref{fig12}. This blob is formed at 133.103 min and located at ($-$22, 16) Mm. Panel (b) displays the time evolution of the number density along the field line where the blob is located, which clearly shows the movement of this blob. The change of the $x$-velocity component and the $y$-velocity component of this blob from 133.103 min (panel (a1)) to 186.058 min (panel (a6)) are denoted by the blue and red lines, respectively. Panel (c) displays the time evolution of the pressure along this field line, and the blue curve shows the trajectory of this blob. The trajectory and the velocities are obtained from the tracking method mentioned before, which is consistent with the slice plot along the field line. The blob first moves upwards and then downwards, and from the pressure change in panel (c), we can see that the change of the movement direction of this blob is closely related to the change of the pressure gradient on the upper and lower sides of the blob. At first, the pressure below the blob is higher than the pressure above it, the blob moves upwards lifted by the pressure gradient, then the blob moves downwards as the pressure ahead becomes bigger, but it immediately changes direction as the pressure gradient changes direction again. The change of the pressure gradient may result from a piston effect due to the compression of the moving blob and the upflows caused by the heating from the footpoints \citep{Adrover2021}. After 174.609 min (indicated by the green dashed line in panel (b)), the blob starts to accelerate downwards and catches up with the condensation next to it, after which they merge and finally fall down to the chromosphere (see panels (a7) and (a8)). 

\begin{figure}
\includegraphics[trim = 0mm 0mm 0mm 10mm, clip, width=1.0\textwidth]{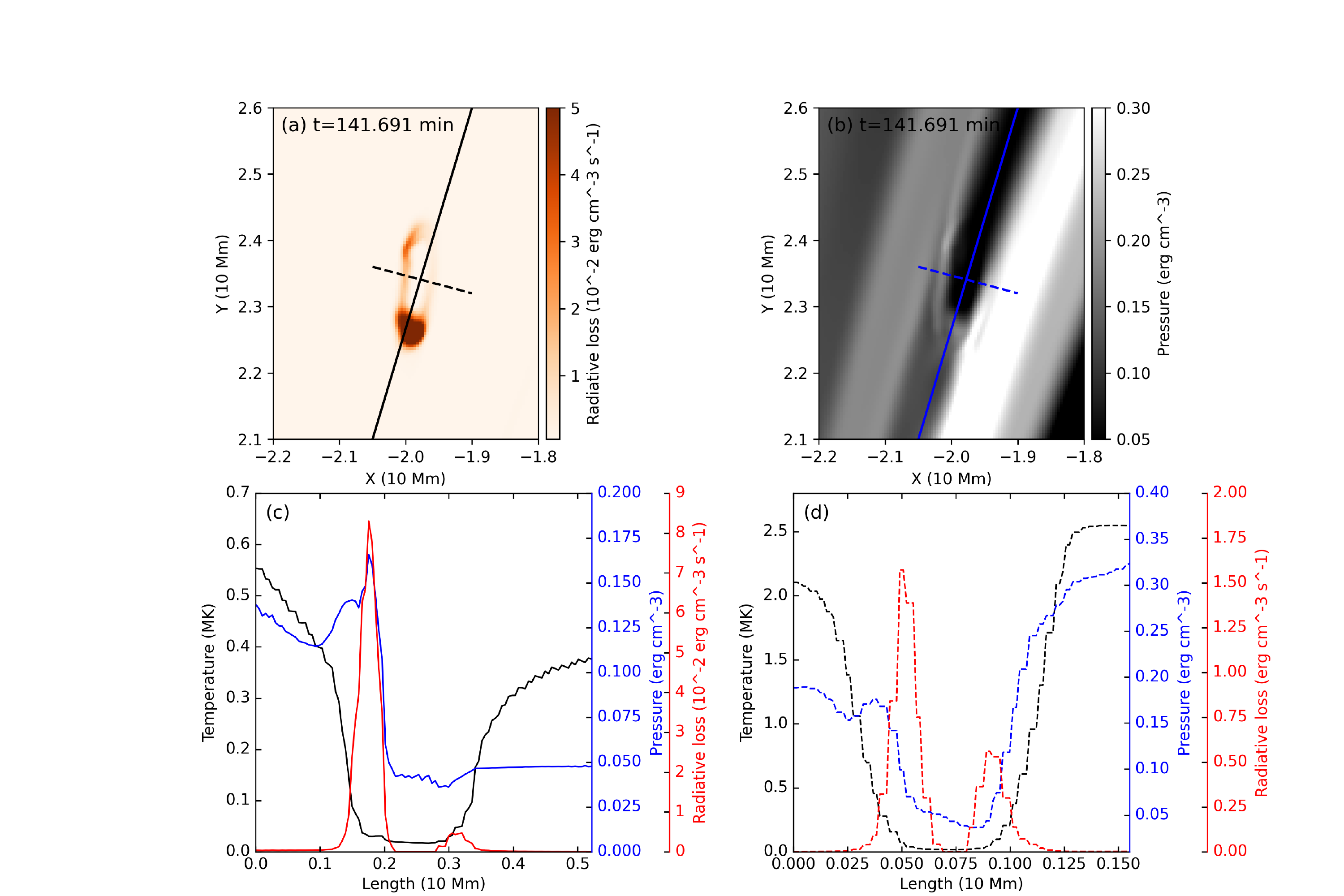}
\caption{The upper panels are the radiative cooling and gas pressure maps at $t$ = 141.691 min for the blob also analyzed in Figure~\ref{fig12}. The lower panels plot temperature (black curves), gas pressure (blue curves), and radiation loss (red curves) along the selected solid and dashed lines crossing the blob center. Panel(c) is along the parallel cut (solid line) while panel (d) is along the perpendicular cut (dashed line).
\label{fig13}}
\end{figure}

Figure~\ref{fig13} shows substructures of this coronal rain blob at 141.691 min, and its field-of-view (FOV) is outlined by the green rectangle in Figure~\ref{fig12}(a2). As pointed by the black dashed vertical line in Figure~\ref{fig12}(b), the blob is moving downwards at this moment, while at the next snapshot it is moving upwards. To better quantify this, we make two lines crossing the center of the blob shown in panels (a) and (b), and the temperature, gas pressure, and radiative loss along these lines are displayed in panels (c) and (d). The solid line is parallel to the field line, while the dashed line is perpendicular to it. As mentioned in the last paragraph, along the field line the pressure below the blob is much higher than the pressure above it (see the blue curve in panel (c)), as the descending blob and the evaporation at the loop footpoint may compress the plasma between them. In agreement with, e.g. \citet{Oliver2014}, the strong pressure gradient supports the blob and slows down the blob in its descent, and even lifts the blob to move upwards, which explains why the blob changes the direction of movement at the next snapshot. The coronal rain blob is surrounded by an area where the temperature declines from more than 0.5 MK outside the blob to 0.02 MK inside the blob. This area is usually considered as PCTR structure in prominences and has been found in coronal rain as well \citep{Fang2015, Antolin2015, Antolin2020}. The radiative loss curves exhibit two peaks near this area, indicating that catastrophic cooling mainly occurs at the boundary of blobs, ensuring the condensation continues to grow. 

\begin{figure}
\centering
\includegraphics[trim = 0mm 10mm 0mm 30mm, clip, width=0.8\textwidth]{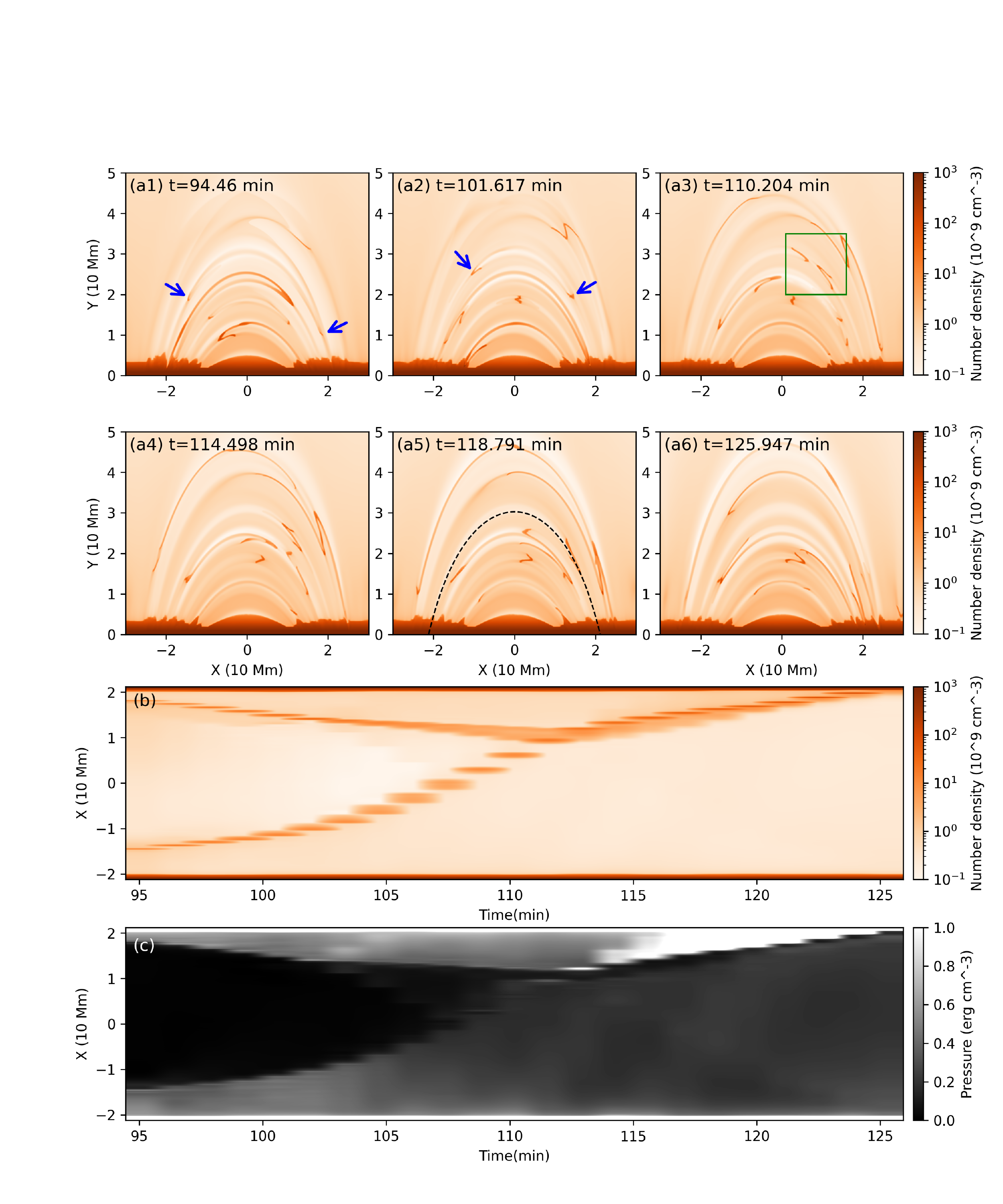}
\caption{Number density map time series (panel (a)), the density evolution (panel (b)) and pressure change (panel (c)) along one magnetic field line showing the movement of two approaching blobs. The blue arrows in panels (a1) and (a2) indicate the two blobs we focus on. The green box in panel (a3) outlines the FOV of Figures~\ref{fig15}(a) and~\ref{fig15}(b), and the black dashed curve in panel (a5) shows the magnetic field line where the two blobs are located.
\label{fig14}}
\end{figure}

\begin{figure}
\includegraphics[trim = 0mm 0mm 0mm 0mm, clip, width=1.0\textwidth]{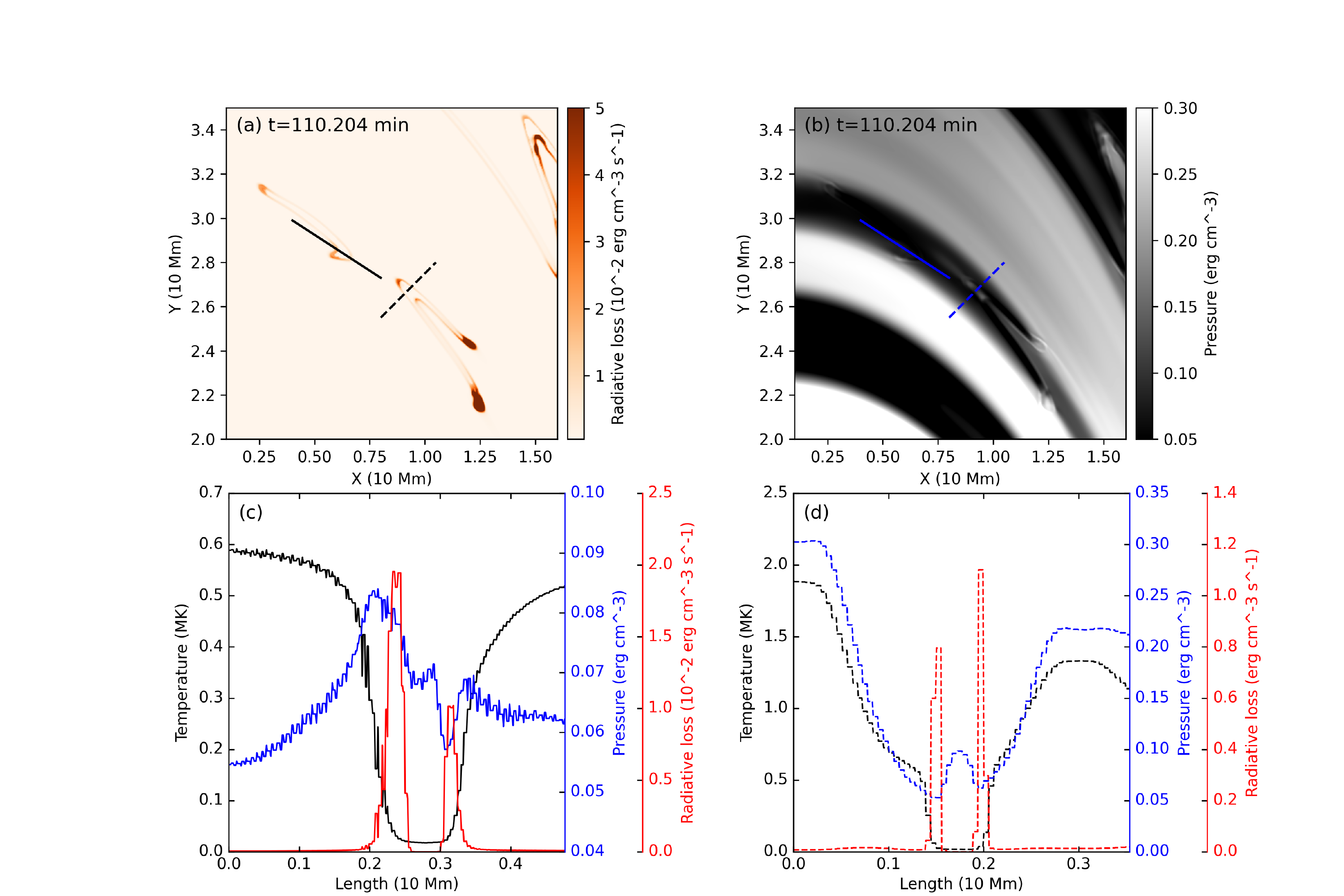}
\caption{The upper panel are the radiative cooling and gas pressure maps at $t$ = 110.204 min for the two mutually approaching blobs in the inset of Fig.~\ref{fig14}. The lower panels plot temperature (black curves), gas pressure (blue curves), and radiation loss (red curves) along the selected solid and dashed lines crossing the blob center. Panel(c) is along the parallel cut (solid line) while panel (d) is along the perpendicular cut (dashed line).
\label{fig15}}
\end{figure}

Except for simple O-shaped blobs, there are many condensations which show V-like structures during propagation. Figures~\ref{fig14}(a1) $-$ (a6) is the number density map showing the movements of blobs during 94.460 min to 125.947 min. The blue arrows in panels (a1) and (a2) indicate two blobs we focus on, which are formed at the same time and located at nearly the same magnetic loops. We choose a magnetic field line passing these two blobs, as denoted by the dashed curve in panel (a5), and the number density and pressure changes along this field line are presented in panels (b) and (c), respectively. These blobs move upwards and approach each other, as the gas pressure between them is much lower than the pressure outside (see panel (c)), which is caused by catastrophic cooling and the lower density in between them (see panel (b)). During this period, the condensations extend from the position where the first condensation occurs, forming larger blobs shown in panel (a2). As the condensations at different loops are experiencing different pressure gradients and gravity, the velocity of the condensation segments are different, and the blobs change their shapes accordingly. Both blobs have faster siphon flows at the position closer to their center, consequently they become deformed and exhibit V-shaped patterns shown in panel (a3). At 114.498 min, these two blobs meet and merge, then they move towards the right footpoints of the coronal loops together and enter the chromosphere ultimately. After they merge together, the density of the new blob becomes bigger (see panel (b)), and it moves downwards against the pressure gradient, as the increased pressure underneath it can't balance out the gravity anymore. The radiative cooling and pressure maps at the green square in panel (a3) are displayed in Figures~\ref{fig15}(a) and~\ref{fig15}(b). The temperature, gas pressure, and radiative loss along the solid and dashed lines are plotted in Figures~\ref{fig15}(c) and~\ref{fig15}(d), respectively. We can see that the V-shaped blobs also have the PCTR area around them. In this area, the temperature changes rapidly from a coronal temperature to a chromospheric temperature, the radiative cooling is stronger and the pressure is lower. During the propagation of these blobs, catastrophic cooling further happens in their heads and tails, resulting in the elongation of the blobs. We can see a higher radiative loss at the heads or tails of both O-shaped and V-shaped blobs in Figures~\ref{fig13} and~\ref{fig15}.

\section{Conclusion and Discussion} \label{sec:conc}

We follow the setup in previous 2.5D simulations of coronal rain, and extend to a more realistic heating pattern where a power-law distribution of Gaussian heating events centered in both space and time (with typical 5-minute durations) is adopted. The formation process and the evolution of coronal rain are reproduced in our simulation. Due to thermal instability, the temperature and gas pressure in the perturbed area decline significantly, therefore ambient plasma is inhaled from the surroundings, and condensations continue to form. In the previous works with similar setups \citep{Fang2013, Fang2015} where the heating was uniform and steady, the first condensation always occurs after a longer time (more than 100 minutes). Here the first condensation starts after 53 minutes, much earlier than the former works, thanks to the TRAC method we adopted, which satisfactorily corrects the underestimated chromospheric evaporation. The fast siphon inflows generate fan-shaped rebound shocks, and the plasma near the shock fronts is heated. The initial asymmetric situations where condensations form bring about different and complicated evolutions of the rebound shocks, and result in asymmetric expansion shock fronts. As the blobs eventually hit the TR and enter the chromosphere, they trigger concurrent upflows which follow the same loops as the downflows. Groups of condensations that occur simultaneously in neighboring magnetic strands are seen as dark falling structures in AIA synthetic hot channels, and bright rain clumps which have a cool core in the cool 304 Å channel. These clumps, with widths up to a few Mm, are known as ``coronal rain showers" \citep{Schrijver2001, Kamio2011, Antolin2012}. Here we show synthetic views on our coronal rain shower events and simulated for over 600 minutes.

We investigate the properties of all the 6060 blobs occurring during our simulation. The statistical analysis of the coronal rain blobs notices that the number and mass seem to have continuous cycles, and the cool plasma mass FFT result shows periodicities of 83 minutes and longer ones. Compared to previous simulation work by \citet{Fang2015}, which showed that a secondary cycle started 50 minutes after the ending of the first one, here the coronal rain continuously occurs during the whole simulation. As mentioned above, this may be a direct result of the randomized heating pattern we adopted. Since the heating is randomized, the chromospheric evaporation rate has different spatial and temporal distributions. Thus, different magnetic loops are going through repeating heating-condensation cycles with different periodicities and at different times. For the whole system, it is heated as soon as the localized heating is turned on, about 53 minutes later radiative cooling takes place and leads to plasma condensation, which is then followed by long competing phases during which the heating and cooling keep happening and the condensations occur all the time. By analyzing the time evolution of the temperature and number density along different magnetic field lines, we found that these field lines have heating $-$ condensation cycles with similar periods, which is resulting from a synchronizing mechanism that coronal rain seems to occur near-simultaneously in neighboring field lines \citep{Fang2013, Fang2015, Antolin2020}. The intensity of the synthetic AIA 94 {\AA},  171 {\AA}, and 193 {\AA} emissions in the corona exhibit periodic cycles similar to the mass of hot plasma, and 304 {\AA} flux curve presents similar morphological changes as the mass of coronal rain. All these flux evolutions have periods of tens of minutes to several hours, which is in agreement with the EUV intensity pulsations \citep{Auchere2014, Auchere2016, Froment2015, Froment2017, Froment2018}. Our results confirm that the coronal rain and the EUV pulsations both correspond to recurrent heating $-$ cooling cycles in the corona, and they may occur together \citep{Froment2020}.

Statistical studies show that most blobs have widths less than 800 km, and lengths from a few hundred km to 25 Mm. Smaller blobs with areas under 0.5 Mm$^2$ dominate the population. The blobs have a broad velocity distribution varying from only a few km s$^{-1}$ to over 120 km s$^{-1}$, which is in accordance with the observations \citep{Muller2005, Antolin2012}. The mean speed of all blobs is about 22.4 km s$^{-1}$, and the acceleration of the majority of blobs is between $-$200 and 200 m s$^{-2}$. We count the proportion of downward moving blobs at each moment and find that upward-moving blobs exist at basically every moment, sometimes even more than the number of falling blobs. The upward motion of coronal rain has been observed, but there is no statistical analysis of the percentage of upward blobs, as it is hard to distinguish the movement of coronal rain blobs due to insufficient resolution. Based on the result, we speculate that there may be more than 20\% of blobs moving upwards in future high temporal and spatial resolution observations. 

Analyzing the dynamics of coronal blobs, we find sub-ballistic fall accelerations and upward motions of the coronal rain blobs. As proposed by \citet{Schrijver2001}, the condensation speeds are determined by the developing pressure gradient in the loops, rather than gravity. \citet{Antolin2010} also suggested that local changes of internal pressure in the loop arising from catastrophic cooling have a greater influence on the dynamics of the condensations than gravity. More and more simulation results support that the gas pressure gradient in the magnetic loop is the primary candidate to explain the descent slower than free-fall and the upward motion of the coronal rain blobs \citep{Fang2013, Oliver2014, Martinez2020}. Apart from gas pressure, the ponderomotive force from transverse MHD waves has also been proposed as a possible mechanism to slow down coronal rain, but it is not expected to play the major role \citep{Antolin2011, Kohutova2016, Verwichte2017}. Here we also find that the strong gas pressure gradient above and below the blobs serves as a levitating force against gravity, influencing the dynamics of the coronal rain. There are several possible mechanisms for the source of developing pressure gradient in the magnetic field line. First of all, the continuous localized heating at the footpoints of the field lines naturally brings about a pressure increase and chromospheric evaporation. Secondly, large pressure variation can be introduced by the thermal instability or catastrophic cooling happening in the field lines. As shown by Figure~\ref{fig14}, two condensations form simultaneously at one field line, and the low pressure in between them is introduced due to catastrophic cooling, and the gas pressure outside drives them to move upwards and merge into one blob. Also, as the condensation moves along the magnetic field line, it may compress the plasma ahead of it. This effect can be seen in Fig.~\ref{fig12}, as the blob moves upwards, the pressure near the upper side of the blob becomes bigger, and the increase of pressure may be more obvious while falling as both the compression and the heating are working. From the statistical analysis of the velocity component along the gravity with the number densities of all blobs (see Figure~\ref{fig10}(d)), we also find that for denser blobs, the upward motions are relatively rare and slow as they need stronger pressure gradient to resist gravity, which is in accordance with the previous simulation works \citep{Oliver2014, Martinez2020, Adrover2021}.

Since the condensations of different magnetic strands experience different pressure gradients and gravity, the shape of the blob changes according to the different speed distribution, which will be elongated and sometimes assume a V-shaped structure as shown in our simulation. Lots of V-shaped condensations are found in our simulations, which can last for more than 20 minutes. This kind of feature has been noticed in previous simulations. \citet{Fang2015} mentioned that shear flows due to the pressure variation in the corona which accompanies thermal instability may play an important role in the formation of these features. \citet{Martinez2020} find that the V-shape blob develops due to its horizontal variation of density, as the denser parts of the blob require a larger pressure gradient to counterbalance the acceleration of gravity, so the falling velocity of the denser parts of the blob is bigger than the lighter ones. It is worth mentioning that these V-shaped condensations haven't been actually detected in observational studies, in which only the elongation of the condensations are reported, with no deformations at the sides of the blobs \citep{Antolin2012}. One possible explanation for this discrepancy is the insufficient resolution of current instruments. More importantly, the shape of condensations may be quite different when extrapolating our 2D results to the 3D situation, thus future 3D studies will be needed to assess how these shapes persist when relaxing the 2.5D assumptions. \citet{Antolin2020} proposed that with axial symmetry, the V-shaped structure may not be observed, but just lead to an elongation of the blobs.

The blobs are found to have a surrounding PCTR structure \citep{Fang2015, Antolin2015, Antolin2020}, where strong radiation loss peaks exist and temperature decreases sharply from a coronal temperature outside to a chromospheric temperature inside. Regarding the area where the temperature declines from 0.5 MK to 0.02MK as the width of this PCTR structure, we find that the width of the PCTR structure is about 500 km parallel to magnetic field lines and around 200 km perpendicular to field lines. Previous studies on solar prominences found that due to the strong parallel conduction, the PCTR parallel to the magnetic field was wide, while the PCTR perpendicular to the magnetic field should be very narrow since the conduction in this direction is weak \citep{Heinzel2001, Vial2015}. The thermal conduction parallel to the magnetic field is several orders of magnitude larger, so the PCTR in the parallel direction should be significantly thicker. Here the PCTR parallel to the field lines is wider than the PCTR in the perpendicular direction, but there is no difference in magnitude. There may be several possible reasons for the relatively small difference. First of all, the resolution of our simulation may not be able to resolve the PCTR. For example, a single coronal rain blob observed under our resolution may actually consist of several blobs, as a result, the PCTR is naturally thicker. In addition, different shearing speeds of the blobs at different positions or different magnetic field lines may cause the PCTR to be thicker across the vertical magnetic line. Finally, the TRAC method we adopted will artificially widen relatively steep temperature gradients, such as the transition region, so the PCTR may be widened. The new TRACL and TRACB method implemented in the AMRVAC \citep{Zhou2021}, which we will adopt in future works, can avoid this error as they can be confined in height to act only at the usual TR in the lower atmosphere. At the heads or tails of both O-shaped and V-shaped blobs, the radiative loss also has higher values as shown in Figures~\ref{fig13} and~\ref{fig15}. These asymmetries in the PCTR structure should have clear consequences for actual non-LTE radiative transfer computations, which typically idealize the PCTR changeover.

\begin{acknowledgments}
We thank the referee for valuable suggestions. This work was supported by the European Research Council (ERC) under the European Unions Horizon 2020 research and innovation program (grant agreement No. 833251 PROMINENT ERC-ADG 2018). The computational resources and services used in this work were provided by the VSC (Flemish Supercomputer Center), funded by the Research Foundation Flanders (FWO) and the Flemish Government - department EWI. This work was further supported by an FWO project G0B4521N and a joint FWO-NSFC grant G0E9619N and by Internal funds KU Leuven, through the project C14/19/089 TRACESpace.
\end{acknowledgments}

\bibliography{manuscript}{}
\bibliographystyle{aasjournal}

\end{document}